\documentclass[preprint]{aastex6}
\usepackage{booktabs}
\usepackage{textcomp}
\usepackage{amsmath}

\newcommand\RR{{\bf R}}
\newcommand\rr{{\bf r}}

\begin{document}


\title{Equilibrium Tidal Response of Jupiter: Detectability by \textit{Juno} Spacecraft}


\author{Sean M. Wahl\altaffilmark{1} }

\affil{Department of Earth and Planetary Science, University of
California, Berkeley, CA, 94720, USA}

\author{Marzia Parisi\altaffilmark{2}}
\affil{Jet Propulsion Laboratory, California Institute of Technology, Pasadena, CA 91109, USA}

\author{William M. Folkner\altaffilmark{3}}
\affil{Jet Propulsion Laboratory, California Institute of Technology, Pasadena, CA 91109, USA}

\author{William B. Hubbard\altaffilmark{4}}
\affil{Lunar and Planetary Laboratory, The University of
Arizona, Tucson, AZ 85721, USA}

\and

\author{Burkhard Militzer\altaffilmark{5}}
\affil{Department of Earth and Planetary Science; Department of Astronomy, University of
California, Berkeley, CA, 94720, USA}


\altaffiltext{1}{swahl@berkeley.edu}
\altaffiltext{2}{marzia.parisi@jpl.nasa.gov}
\altaffiltext{3}{william.m.folkner@jpl.nasa.gov}
\altaffiltext{4}{hubbard@lpl.arizona.edu}
\altaffiltext{5}{militzer@berkeley.edu}

\begin{abstract}

    An observation of Jupiter's tidal response is anticipated for the on-going
    \textit{Juno} spacecraft mission. We combine self-consistent, numerical
    models of Jupiter's equilibrium tidal response with observed Doppler shifts
    from the \textit{Juno} gravity science experiment to test the sensitivity of
    the spacecraft to tides raised by the Galilean satellites and the Sun. The
    concentric Maclaurin spheroid (CMS) method finds the equilibrium shape and
    gravity field of a rotating, liquid planet with the tide raised by a
    satellite, expanded in Love numbers ($k_{nm}$). We present improvements to
    CMS theory that eliminate an unphysical center of mass offset  and study in
    detail the convergence behavior of the CMS approach.  We demonstrate that
    the dependence of $k_{nm}$ with orbital distance is important when
    considering the combined tidal response for Jupiter. Conversely, the details
    of the interior structure have a negligible influence on $k_{nm}$, for
    models that match the zonal harmonics $J_2$, $J_4$ and $J_6$, already
    measured to high precision by \textit{Juno}.  As the mission continues,
    improved coverage of Jupiter’s gravity field at different phases of Io’s
    orbit is expected to yield an observed value for the degree-2 Love number
    ($k_{22}$) and potentially select higher--degree $k_{nm}$. We present a test
    of the sensitivity of the \textit{Juno} Doppler signal to the calculated
    $k_{nm}$, which suggests the detectability of $k_{33}$, $k_{42}$ and
    $k_{31}$, in addition to $k_{22}$. A mismatch of a robust \textit{Juno}
    observation with the remarkably small range in calculated Io equilibrium
    $k_{22}=0.58976\pm0.0001$, would indicate a heretofore uncharacterized
    dynamic contribution to the tides.

\end{abstract}

\keywords{Jovian planets, Tides, Interiors, Jupiter}



\newcommand{\sphint}{\int_{-1}^1 d\mu' \int_{0}^{2\pi} d\phi' \int_{r' > r} dr'}
\newcommand{\sphshort}{\int_{\tau} d\tau}
\newcommand{\muint}{\int_{-1}^1 d\mu'}
\newcommand{\phiint}{\int_{0}^{2\pi} d\phi'}
\newcommand{\xiint}{\int_{b/a}^{\xi} d\xi'}

\section{Introduction} \label{sec:intro}

Since orbital insertion in July of 2016, the \textit{Juno} spacecraft  has
yielded robust measurements of even \citep{Folkner2017a} and odd
\citep{Iess2018} zonal harmonics  of Jupiter's gravitational field. These
high-precision gravity measurements inform our understanding of Jupiter's
interior structure \citep{Wahl2017a,Nettelmann2017,Debras2019} and wind
structure \citep{Kaspi2017a,Kaspi2018,Guillot2018}. As the spacecraft makes
additional orbits, it is expected  to constrain the tidal response of the planet
in terms of the second degree Love number, $k_{22}$, and possibly higher degree
$k_{nm}$ as well. Prior knowledge of the higher degree $k_{nm}$ from theoretical
models also aids in the fitting of a gravity solution to the \textit{Juno}
Doppler radio measurements.

The concentric Maclaurin spheroid (CMS) method \citep{Hubbard2012,Hubbard2013}
is a non-perturbative, numerical method for calculating the self-consistent
mass distribution and gravity field of a rotating liquid body. In
two-dimensions, the CMS approach allows efficient and precise exploration of
interior structure models \citep{Militzer2019a}. In three-dimensions, the CMS
method permits a precise calculation of equilibrium tidal response, yielding
both $k_{22}$ and higher degree $k_{nm}$ \citep{Wahl2017b,Wahl2016}. The CMS
calculations demonstrated a significant effect from the rotational flattening of
Jupiter and Saturn, manifesting in an enhanced $k_{22}$ compared with
calculations treating the tidal response as a perturbation from a sphere
\citep{Gavrilov1977}, as well as splitting of the higher degree $k_{nm}$. In
this paper we present improvements to the CMS equilibrium tidal response
calculations, including a solution to the unphysical center of mass offsets
described in \citet{Wahl2017b}. We also study in greater detail the convergence
behavior of CMS tidal response calculations, and how this affects the precision
to which we can determine different values of $k_{nm}$ for the four Galilean
satellites, and the Sun.

A preliminary value of $k_{22}=0.625\pm0.06$ was presented in the \textit{Juno}
gravity solution of \citet{Iess2018}, consistent with the enhanced value
predicted from CMS, but not sufficiently precise to fully characterize the
nature of the tidal response.  In the case of Saturn, astrometric observations
\citep{Lainey2017} yield a $k_{22}$ that is below the value for a equilibrium
tide to their reported uncertainty, but still enhanced compared to the
prediction from perturbation theory applied to a spherical Saturn.

A major outstanding question is whether Jupiter's tidal response can be
adequately described by equilibrium theory alone, as addressed in this paper, or
whether there is a detectable dynamic contribution to the tides. The equilibrium
tidal response treated here assumes that the rotating liquid interior responds
instantaneously to the perturbing satellite\footnote{  The terms ``equilibrium
tides'' and ``static'' tides have been used interchangeably to refer to the same
phenomenon in the literature. We elect to use ``equilibrium'' in this paper}. In
the co-rotating frame of the planet, we find the equilibrium tidal distortion,
fully determined by the orbit and $k_{nm}$ calculated for each perturbing
satellite. In contrast, dynamic tides are a frequency dependent response,
affected by proximity to resonances and excitation and waves as the tidal
perturbations interact with interior structural features, such as stable layers
or density discontinuities \citep{Marley1993,Fuller2014,Fuller2014b}.
Calculations of dynamic tides rely on normal mode calculations
\citep{Durante2017}, related to techniques used in astroseismology
\cite{Chaplin2013}. While special cases of dynamic tidal response been
demonstrated to result in a small enhancement to $k_{22}$ \citep{Gavrilov1977},
a comprehensive treatment of dynamic  tidal response of a Jovian planet with
rotational flattening has not been attempted. However, once an interior model is
fitted to observed zonal harmonic coefficients, $J_n$, its equilibrium tidal
response coefficients, $k_{nm}$, are determined with little remaining
uncertainty. Thus, measurements of $k_{nm}$ that disagree with the equilibrium
response are evidence of a dynamic tidal response.

In \citet{Wahl2016}, we presented preliminary calculations of Jupiter's
equilibrium tidal response for a range of different interior models
\citep{hubbardMilitzer2016} based on pre-\textit{Juno} knowledge and assumptions
about the planet's interior. This included a wide range of models, in part
because of a disagreement of the JUP310 gravity solution \citep{Jacobson2013}
with earlier gravity solutions \citep{Campbell1985,Jacobson2003}. The
pre-\textit{Juno} gravity solution permitted two or three-layer interior
structure models with a variety of hydrogen-helium equations of state
\citep{Saumon2004,Militzer2008,Nettelmann2012,Miguel2016,hubbardMilitzer2016}.
As of the \textit{Juno} gravity solution presented by \citet{Iess2018}, there
are now precise determinations of the even gravity harmonics up to $J_{10}$ and
odd harmonics up to $J_{9}$. The low-order, even gravity harmonics $J_2$, $J_4$
and $J_6$, are all determined to $\sim10^{-8}$ or better. When taken into
consideration with state of the art equations of state for hydrogen-helium
mixtures \citep{Vorberger2007,Militzer2013} and constraints on atmospheric
composition and temperature, these low-order harmonics place strong constraints
on the viable interior structure models
\citep{Wahl2017b,Nettelmann2017,Debras2019,Militzer2019b}, suggesting that
Jupiter's interior structure is more complicated the previously thought.
\citet{Nettelmann2019} presented calculations of equilibrium $k_{nm}$ of an
interior structural fit to earlier \textit{Juno} gravity solution
\citep{Folkner2017a}. In this paper we present results for the equilibrium
tidal response of Jupiter, with the constraints from the most recent
\textit{Juno} gravity solution \citep{Iess2018} and an improved implementation
of the CMS method. We consider the ranges in tidal response expected from
uncertainty from the interior structure, deep winds and measurements of  the
relevant physical parameters, as well as the numerical precision of CMS method.
Characterizing this uncertainty in modeling the equilibrium tide is necessary to
know whether a precise measurement of $k_{nm}$ alone is sufficient to detect a
dynamic contribution, and distinguish it from the equilibrium contributions.

In Section \ref{sec:cms} we present an overview of the CMS method focused on the
details pertinent to the calculation of the equilibrium tidal response. In
Section \ref{sec:update} we describe the improvements to the CMS approach for
equilibrium tidal calculations, and their effect  on the calculations. In
\ref{sec:interior} we describe interior  models used in this study, as well as
our methods for quantifying the possible range of  $k_{nm}$ from interior
structure and winds. In Section \ref{sec:convergence} we look at the convergence
of the CMS models with various parameters to quantify precision to which
different $k_{nm}$ can be predicted. In Section \ref{sec:galilean} we present,
in detail, the tidal response of Jupiter to the four Galilean satellites and the
Sun, and consider  \textit{Juno}'s ability to sample that tidal response.
Finally in Section \ref{sec:doppler}, we test the sensitivity of the
\textit{Juno} Doppler  gravity measurements to the calculated $k_{nm}$, and
discuss the significance of the higher degree $k_{nm}$ to the fitting of a
gravity solution by the \textit{Juno} gravity science experiment.

\section{Methods} \label{sec:methods}

\subsection{CMS method} \label{sec:cms}

The \textit{concentric Maclaurin spheroid} method (CMS) is a non-perturbative,
iterative method to find the gravity field of a liquid body, which was
formulated by \citet{Hubbard2012,Hubbard2013} and extended to three dimensions
by \citet{Wahl2017b}.

\begin{figure}[h!]
  \centering
    \includegraphics[width=0.7\textwidth]{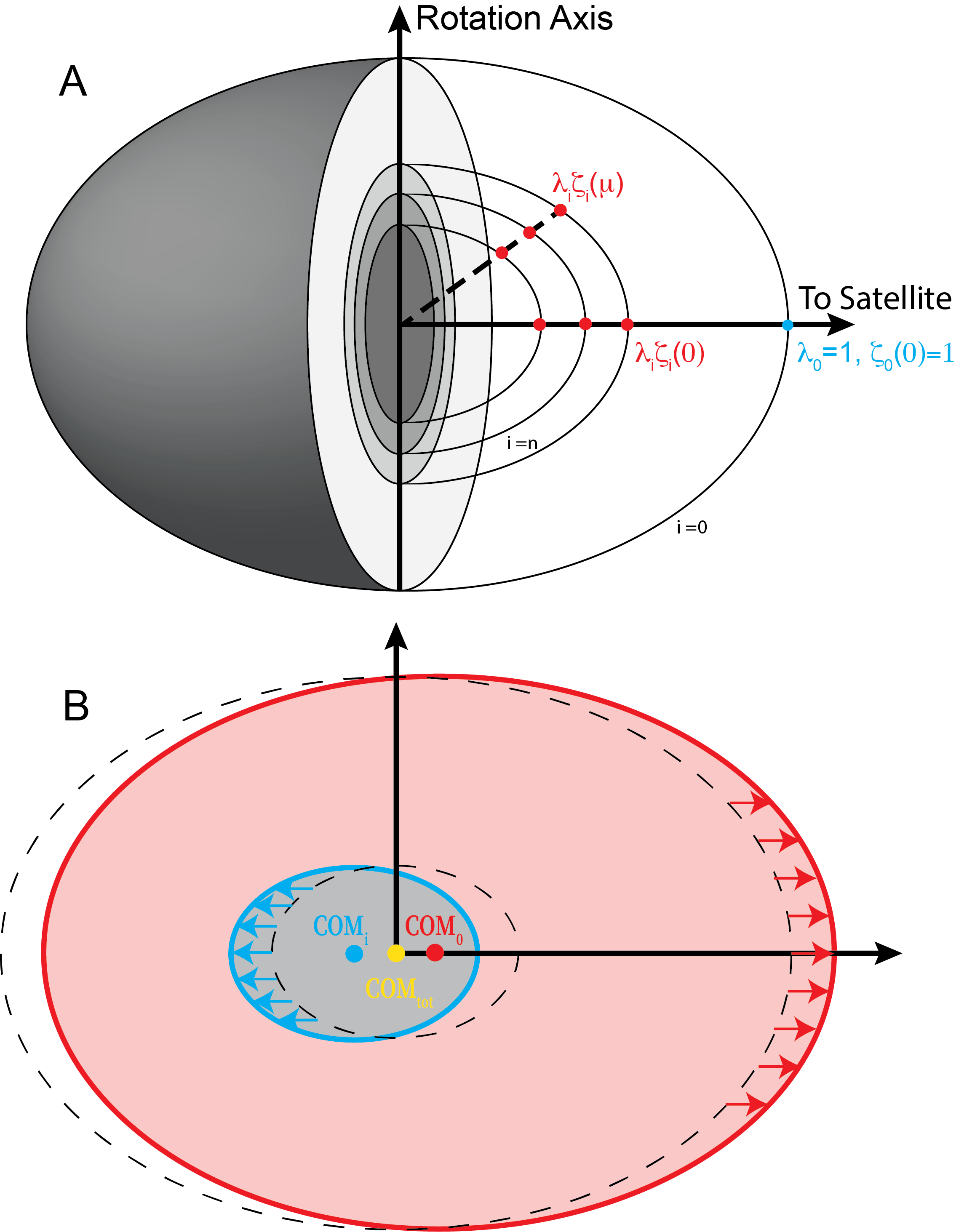}
\caption{ Conceptual diagram of a Concentric Maclaurin Spheroid (CMS) model with
a tidal perturbation from a satellite. A: The density structure is discretized
into a superposition of constant density spheroids. The surface of spheroid i is
described by the fraction $\zeta_i$ of the equatorial radius $\lambda_i$, at
coordinates $\mu$, $\phi$. B: In the presence of a tidal perturber the spheroids
surfaces and centers of mass (COM) are permitted to respond to the external
potential under the constraint of constant volume, such that the outermost
spheroids shift towards the satellite while the innermost spheroids shift away
from the satellite. The total center of mass remains fixed at the origin. }
\label{fig:spheroid}
\end{figure}

In this method, a continuous density structure is discretized into $N$ nested,
constant-density spheroids, as shown in Fig. \ref{fig:spheroid}. The planet's
self-gravity, $V$,  is calculated as a volume-integrated function of all
spheroids in their current configuration, and combined with a centrifugal
potential, $Q$, and an external potential from a perturbing satellite, $W$, into
a single effective potential,
\begin{equation} U({\bf r}) = V({\bf r}) + Q({\bf r}) + W({\bf r}) \label{eq:effective_potential} \end{equation}
The shape of each spheroid is then adjusted until the surface of each
becomes an equipotential surface of $U$.

Given a prescribed interior density structure, the non-spherical contributions
to the potential are parameterized by three non-dimensional numbers. First is
the relative strength of the centrifugal potential,
\begin{equation} q_{\rm rot} = \frac{\omega^2 r_{\rm J}^3}{GM_{\rm J}}, \label{eq:qrot} \end{equation}
where $\omega$ is the sidereal rotation frequency, $r_{\rm J}$ is the equatorial
radius of the Jupiter, $G$ is the universal gravitational constant and $M_{\rm
J}$ is the mass of Jupiter. The second describes the relative strength of the
tidal perturbation,
\begin{equation} q_{\rm tid} = -\frac{3m_{\rm s}r_{\rm J}^3}{M_{\rm J}R^3}, \label{eq:qtid}
\end{equation}
where $m_{\rm s}$ is the mass of the perturbing satellite (or the Sun) and $R$ is
the orbital distance. Last is the ratio of the satellite's orbital distance to
the planet's radius,
\begin{equation} R/r_{J}. \label{eq:radius_ratio}
\end{equation}
The relevant physical parameters for Jupiter, its satellites and the Sun are
summarized in Tab. \ref{tab:params}. The uncertainty of $q_{\rm rot}$ is
dominated by the $\sim$4 km uncertainty in $r_J$. It is worth noting that although
a CMS calculation is performed for a single set of parameters, $q_{\rm tid}$ and
$R/r_{J}$ vary with time due to the eccentricity of the orbit, which must be
taken into account when translating the calculated $k_{nm}$ into a gravity
signal.

\begin{figure}[h!]
  \centering
    \includegraphics[width=0.7\textwidth]{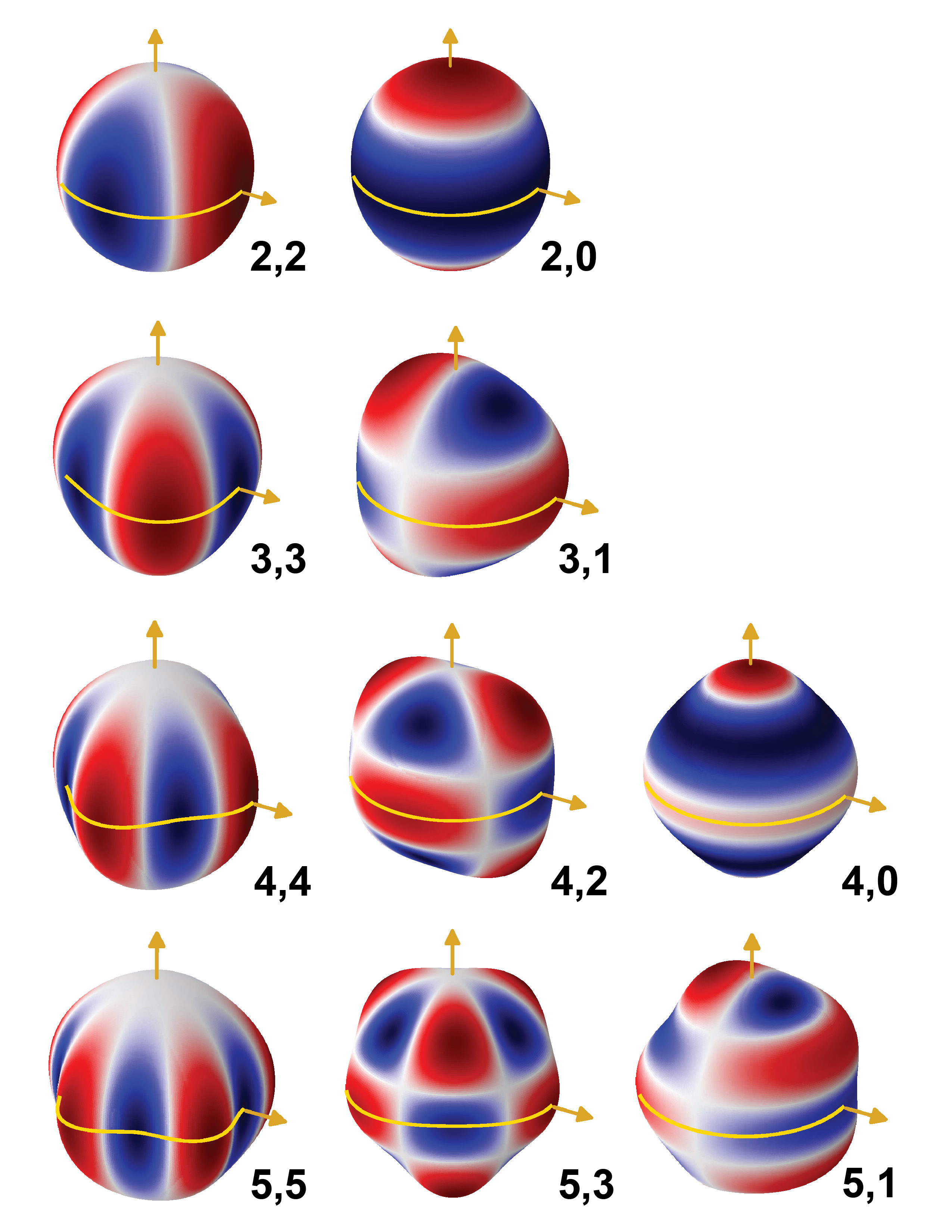}
\caption{ Shapes of the gravity harmonics, $C_{nm}$, labeled by degree $n$ and
order $m$, where $|m|\le n$. Yellow arrows show the pole and indicate the
direction of the tidal perturber, and yellow line shows the equator. }
\label{fig:harmonics}
\end{figure}

The zonal, $J_n$, and tesseral, $C_{nm}$ and $S_{nm}$, harmonics of the gravity
field (depicted in Fig. \ref{fig:harmonics}) can be calculated by integrating
over the density and shape of a converged CMS model \citep{Wahl2017b}. For
Jupiter and its Galilean satellites $q_{\rm tid} \ll q_{\rm rot}$, which means
that the equilibrium tidal response from each satellite can be calculated
independently, and the resulting $C_{nm}$ and $S_{nm}$ can be linearly
superimposed to obtain the total equilibrium tidal response. The calculations
can be further simplified by assuming that the satellites reside in the planet's
equatorial plane since the Galilean moons each exhibit a small orbital
inclination. Under these assumptions, we perform CMS calculations for a single
satellite at a time, with a fixed position in the equatorial plane.

As per convention \citep{zharkov1978}, the tidal Love numbers
represent ratio of the tidally induced gravity moments to the strength of perturbing
tidal potential, which can be represented as
\begin{equation}
    k_{nm} = -\frac{3}{2}\frac{(n+m)!}{(n-m)!}\frac{(C_{nm}-C_{nm}^0)}{P_n^m(0)q_{\rm tid}}
    \left( \frac{r_{\rm eq}}{R} \right)^{2-n},
\label{eq:kn}
\end{equation}
where $m\le n$ and $P_n^m$ are the associated Legendre polynomials\footnote{This
equation appeared in \citet{Wahl2017b} with a typographical error in the
prefactor}. Here $C_{nm}^0$ is the harmonic for the unperturbed body (i.e. with
rotation, but without tides). For $m\ne0$, $C_{nm}^0=0$, but with rotation
\begin{equation}
    C_{n0}^0 = -2 J_n^0,
\label{eq:cn0_jn}
\end{equation}
where $J_n$ are the corresponding zonal harmonics from an axisymmetric
calculation with the same $q_{\rm rot}$.

For this reason the $k_{n0}$ cannot be directly measured in the \textit{Juno}
Doppler data, as their contribution is indistinguishable from contributions from
rotation and interior density distribution. The eccentricity of the satellite
orbit induces  a small, time-variable component of $C_{n0}$, that might be
detectable as a variation  of the observed $J_n$ with the satellites orbital
phase. If the tidal perturbers are located on the equatorial plane, as we
assume, then all $k_{nm}$ with odd values of $n-m$ are zero. For a non-rotating
planet, the $k_{nm}$ with the same degree $n$ are degenerate with order $m$.
Jupiter, on the other hand, exhibits significant splitting of these $k_{nm}$ due
to the significant rotational flattening.

\subsection{Tidal Response Calculations} \label{sec:update}

In this section, we provide details to perform accurate tidal response
calculation with the CMS method and discuss a number of assumptions and
approximations. Most importantly, we only deal with the {\it equilibrium}
response and assume the planet responds instantaneously to an external
perturbation by a satellite. Since the satellite's diameter is small compared to
its orbital distance, $\bf R$, it is well-justified to treat the satellite as a
point mass, $m_s$. Its gravitational potential is given by
\begin{equation}
  W(\rr,\RR) = \frac{Gm_s}{\left|\RR - \rr \right|},
\end{equation} \label{eq:direct_potential}
and can be expressed in terms of Legendre polynomials,
\begin{equation} \begin{aligned} W(r,\mu,\phi) =& \frac{Gm_{\rm s}}{R}
        \sum_{n=0}^\infty \left( \frac{r}{R} \right)^n \left[ P_n(\mu) P_n(\mu_s) \phantom{\frac{1}{1}}\right. \\
        & \left. + 2 \sum_{m=1}^{n} \frac{(n-m)!}{(n+m)!} \cos(m\phi - m\phi_s)
    P_n^m(\mu) P_n^m(\mu_s) \right].  \end{aligned}
\label{eq:tidal_general}
\end{equation}
where we have specified the satellites location with $R$, $\mu_s$, and $\phi_s$.

For potential theory to be applicable, the tidal perturbation needs to be time
independent, which means one can only derive the tidal response for an analogue
system where the satellite's orbital period is set equal to the rotation period
of the planet. In the rotating frame, the satellite's gravity field then becomes
time independent and it is a well-posed but simplified task to determine the
planet's response.  There are situations where these assumptions are well
justified, e.g., tidally locked exoplanets that have equal orbital and rotation
periods. However, satellites in the solar system all have orbital periods that
are much longer than the rotation periods of the host planets. This introduces a
time dependence into the tidal perturbation and may lead to {\it dynamic} tidal
effects. The dynamic response is typically studied by expanding the planet's
response in terms of normal modes~\cite{Gavrilov1977}. Even when such dynamic
tidal calculations are performed for Jupiter, one expects to find a negligible
tidal lag because the viscosity in giant planets is very small. A counter
example are the tides on Earth, where there is a more substantial response lag
for the solid mantle and crust.

The standard approach to derive a time independent solution is to remove the
average force that the tidal perturber exerts on the planet~\citep{Murray1999}.
Removing the average force can be motivated by representing the planet by a
system of $N$ fluid parcels of mass $m_i$ at locations $\rr_i$. Its total energy
is assumed to be given by $ \mathcal{V}(\rr_1 \ldots \rr_N)$ and an equilibrium
configuration must satisfy ${\bf F}_i = \partial \mathcal{V} / \partial
\rr_i=0$.  In order to establish the orbital distance for a given planet, we
constrain the planet's center of mass to reside at $\RR_{\rm CM}$. To solve this
constrained optimization problem, we introduce the modified function,
\begin{equation}
\tilde{\mathcal{V}}(\rr_1 \ldots \rr_N) = \mathcal{V}(\rr_1 \ldots \rr_N) - \lambda \left[ \sum_i m_i \rr_i - M \RR_{\rm CM} \right]\;\;,
\end{equation}
where $\lambda$ is a Lagrange multiplier and $M$ is the total mass. Solving
$\partial \tilde{\mathcal{V}} / \partial \rr_i=0$ yields that $\lambda$ must be
equal to the average force, $\left< {\bf F} \right>$. The equilibrium solution
of the constrained system must then satisfy, $0 = \partial \mathcal{V} / \partial
\rr_i - \left< {\bf F} \right> \, m_i$, which explains why one would want to
remove the average force.

In the CMS calculations, we derive this average force as follows,
\begin{equation}
\left< {\bf F} \right> = \frac{1}{M} \int d\rr \, \rho(\rr) \, \nabla_{\rr} W(\rr,\RR) = - \frac{1}{M} \nabla_{\RR} \int d\rr \, \rho(\rr) \, W(\rr,\RR) = - \frac{m_s}{M} \nabla_{\RR} V(\rr=\RR)\;\;,
\label{eq:force0}
\end{equation}
where  we have integrated over all fluid parcels in the planet in the first two
terms. In the last term, we have used the symmetry of the gravitational
potential, which implies that the average force that the satellite exerts on the
planet is equal but opposite to the force that the planet exerts on the
satellite. The evaluation of the last term is straightforward within the CMS
method, because the gradient is already needed to converge the spheroid shapes
onto equipotential surfaces using Newton's method.

With the average force, we define a modified tidal potential
\begin{equation}
\tilde W(\rr,\RR) = W(\rr,\RR) - \left< {\bf F} \right> \cdot \rr
\label{eq:av_force}
\end{equation}
and introduce it into Eq.~\ref{eq:effective_potential} before using the modified
potential to construct equipotential surfaces. With this approach, one obtains a
stable numerical algorithm that converges to a self-consistent CMS solution.
This algorithm is used for all the results reported in this article.

With other, more approximate approaches, it is more difficult to reach well
converged simulations. \citet{zharkov1978} did not derive the average force
explicitly but instead removed the $n=1$ term from Eq.~\ref{eq:tidal_general},
which is linear in $r$. For point masses, this is equivalent to
Eq.~\ref{eq:force0} but for an extended planet this introduces a small spurious
force that lets the planet drift towards the satellite because the gravitational
force is nonlinear. In the previous implementation of the CMS tidal response
calculation~\citep{Wahl2017b} that was based on Eq.~\ref{eq:tidal_general}, we
noticed a small but persistent center of mass shift ($C_{11}>0$) following each
iterative update of the spheroid surfaces ($\zeta_i$). This was accounted for by
applying a translation to all grid points after each iteration to eliminate
$C_{11}$, but this prevented full convergence of the spheroids to equipotential
surfaces with the expected numerical precision. With our new approach based on
Eqs.~\ref{eq:force0} and \ref{eq:av_force}, this problem has been eliminated and
we now obtained well-converged equipotentials and the computed $C_{11}$ is zero
to within numerical precision.

We initialize our tidal calculations with spheroid shapes defined by the
fractional radius, $\zeta_{i}(\mu,\phi)$, that we obtain from a fully converged
2D axisymmetric CMS solution, $\zeta_{i}(\mu)$. The spheroid shapes have
converged when $U(\zeta_{i}(\mu,\phi))$ is the same for all points in any given
spheroids. We can no longer fix $\zeta_{i}(\mu=0)=1$ since we expect two tidal
bulges to form. Instead we require the volume of each spheroid to be the same as
determined by the initial axisymmetric calculation. Since there is much
flexibility in the 3D CMS calculation, the implementation of the volume
constraint requires some care. For a given spheroid, we first compute a target
potential value, $U^{\rm T}_i$ by averaging $U(\zeta_{i}(\mu_k,\phi_k))$ over
all spheroid points, $k$. We then compute vector of proposed $\zeta$ corrections
with Newton's method,
\begin{equation}
\Delta \zeta_i^{(k)} \equiv \Delta \zeta_i(\mu_k,\phi_k) = \frac { U(\zeta_{i}(\mu_k,\phi_k)) - U^{\rm T}_i } { U'(\zeta_{i}(\mu_k,\phi_k)) }
\end{equation}
where $U'$ is the derivative of the $U$ with respect to $\zeta$. We
require that all $\Delta \zeta_i^{(k)}$ combined do not change the
spheroid volume, $\Omega_i$. We thus remove the volume-changing vector
component with
\begin{equation}
\vec{y}^{\, \rm corr} = \vec{y} - \frac{\vec{y}\cdot \vec{x}}{\vec{x}
  \cdot \vec{x}} \, \vec{x},
\end{equation}
where $\vec{y}$ and $\vec{x}$ represent the vectors $\Delta \zeta_i^{(k)}$ and
$d \Omega_i / d \zeta_i^{(k)}$ respectively. We then apply the corrected
$\vec{y}^{\, \rm corr} \equiv \Delta \zeta_i^{(k)}$ and rescale the spheroid
again to exactly match the original $\Omega_i$.

Once all spheroids have been updated, we compute the new $C_{11}$ term and apply
a single coherent shift to all spheroids, so that planet's total center of mass
is again at the origin. This requires performing a spline interpolation for each
spheroid over $\mu$ and $\phi$, so that the pre and post-shift grid points
remain on the quadrature points. In contrast to the behavior described in
\citet{Wahl2017b}, this center of mass shift gradually decreases to zero as the
spheroids converge towards  equipotential surfaces.

When we examine the converged CMS solutions, we notice that the centers of mass
of individual outer spheroids exhibit a small shift towards the perturbing
satellite, while the inner spheroids have drifted away from it, as shown in the
lower panel of Fig.~\ref{fig:spheroid}. When we instead restricted the center of
mass of every individual spheroids during the CMS iterations, we were not able
to construct equipotential surfaces. We thus conclude that the small spheroid
shifts are necessary to correctly represent how a fluid planet with a realistic
interior density structure responds to the tidal perturbations.

\begin{figure}[h!]
  \centering
    \includegraphics[width=0.45\textwidth]{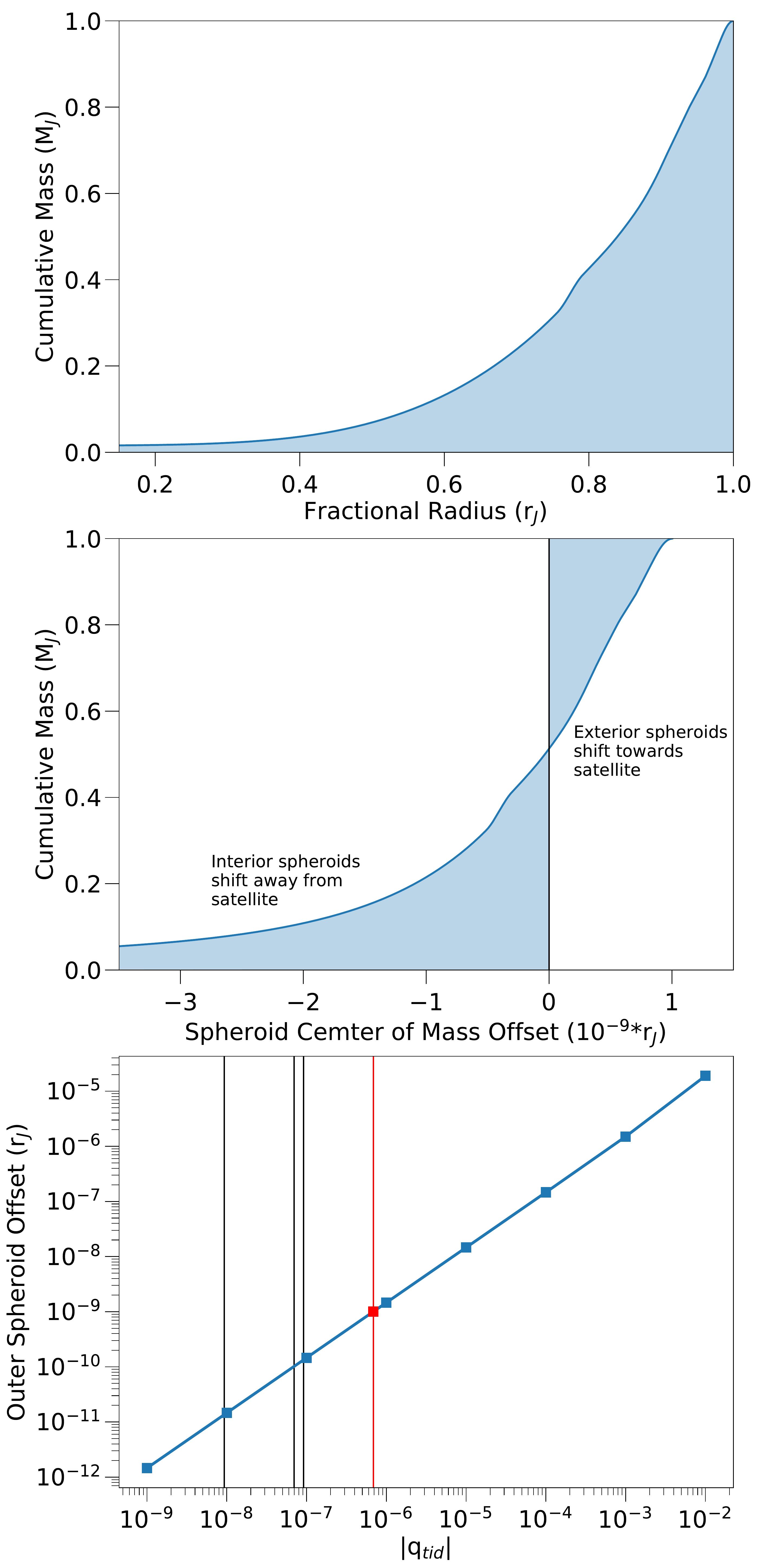}
\caption{ Top panel: Cumulative mass $m/M_J$ as a function of fractional radius
$\lambda$ for a representative interior model. Middle Panel: The offset of
spheroid center of mass as a function of cumulative mass for CMS models with
tidal response to Io. Spheroids exterior to $m/M_J\sim0.5$ have center of mass
(COM) shifts towards the satellite, while interior spheroids have COM shifts
away from the satellite. The total center of mass of the planet is constrained
to lie at the origin. Bottom panel: The magnitude of the COM shift for the
outermost layer ($\lambda=0$) as a function of tidal perturber strength, $q_{\rm
tid}$. Vertical lines denote $q_{\rm tid}$ of the four Galilean moons, with Io
shown in red. }
\label{fig:offset}
\end{figure}

In Fig.~\ref{fig:offset} we show the spheroid shifts quantitatively for a
representative CMS calculation. Over the course of the  calculations the
spheroids arrange themselves so that the centers of mass follow a smooth
function of radius. When shown as a function of integrated mass, it can be seen
that roughly half of the mass is shifted towards the satellite and the  other
half away, such that the planet's total center of mass remains at the origin.
The inner spheroids exhibit a larger magnitude of shift than  the outer
spheroids, but they also contain less mass. Depending on the magnitude of the
tidal perturbation, we find the outermost spheroid to shift between 10$^{-12}$
and 10$^{-5}$ Jupiter radii.

\subsection{Interior models} \label{sec:interior}

We start from the assumption of a liquid planet in hydrostatic equilibrium,
\begin{equation} \nabla P = \rho \nabla U,
    \label{eq:hydrostatic} \end{equation}
where $P$ is the pressure, $\rho$ is the mass density and $U$ the total
effective potential. The material properties of hydrogen--helium mixture, with a
mass fraction  of heavier elements in solution, determines a barotrope $P(\rho)$
for the planet's interior.

Presently, the most trusted equations of states are constructed from \textit{ab
initio} simulations using density functional molecular dynamics (DFT-MD)
\citep{Vorberger2007,Militzer2013,Becker2015,Chabrier2019}.  Our models use
barotropes constructed from a grid of adiabats determined by the
\citet{Militzer2013} equation of state for a hydrogen-helium mixture. The DFT-MD
simulations were performed, with cells containing $N_{He}=18$ helium and
$N_{H}=220$ hydrogen atoms, using the Perdew-Burke-Ernzerhof (PBE) functional
\citep{PBE} in combination with a thermodynamic integration technique.

There was initial disagreement between different DFT-MD based equations of state
for hydrogen helium mixtures, with the REOS \citep{Nettelmann2008,Becker2015}
equation of state predicting hotter, less dense barotropes
\citep{Militzer2009,Guillot2018} than that of \citet{Militzer2013}, due to their
different method for calculating the specific entropy. There is now better
agreement between the independently constructed DFT equations of state
\citep{MH13,Schottler2018,Chabrier2019}, all predicting colder, denser
barotropes consistent with those based on thermodynamic integration.

We follow the same treatment of specific entropy, helium and heavy-element
fraction  as in previous work
\citep{hubbardMilitzer2016,Militzer2016,Wahl2016,Wahl2017a}. The entropy, $S$,
is a proxy for a particular adiabatic temperature $T(P)$ relationship for a
fixed composition H-He mixture ($Y_0=0.245$). This initial composition also
provides the reference barotrope densities, in which deviations from the
baseline composition ($Y_0=0.245$, $Z_0=0$), are treated as perturbations using
the additive volume law \citep{Wahl2017b}.

When the inherent density of the hydrogen--helium mixture is set by the DFT
equation of state, it is not possible to find simple 3-layer interior models
that simultaneously match the \textit{Juno} gravity solution, while satisfying
atmospheric constraints on  temperature and composition from the
\textit{Galileo} entry probe \citep{seiff-1998,vonzahn-jgr-98}. Satisfying all
such constraints requires either more complex interior thermal and compositional
structure \citep{Debras2019}, contributions from deep winds
\citep{Guillot2018,Kaspi2018} or both \citep{Militzer2019b}. The requisite deep
wind profile decays with depth, due to interaction of the conductive fluid with
the magnetic field \citep{Cao2017}, and cannot be described self-consistently
using a potential based theory like CMS \citep{Militzer2019a}.

The Love numbers are known to be strongly correlated with $J_n$ \citep{Wahl2017b}. We
thus elect to first examine a set of simple interior models, capable of matching the
updated low--order axisymmetric \textit{Juno} gravity solution \citep{Iess2018} with
loosened compositional constraints (A1--4 in Tab. \ref{tab:models}). We next consider
the effect of deep wind profiles optimized to match the odd zonal harmonics
\citep{Kaspi2018} (B1--2 in Tab.  \ref{tab:models}). Finally, we include models in
which a more complicated interior structure and deep wind profile are optimized
simultaneously \citep{Militzer2019b} (C1--4 in Tab. \ref{tab:models}).  The models
considering deep winds are described in more detail in Section \ref{sec:winds}.

The first set of models considered are 3 or 4--layer interior models modified from
those presented in \citet{Wahl2017a}, with parameters tuned to match $J_2$, $J_4$ and
$J_6$. These models are denoted as A1--4 in Tab. \ref{tab:models}, with the model
$J_n$ compared to the target values from \textit{Juno}.  The target $J_n$ are
obtained from the \citet{Iess2018} gravity solution to the \textit{Juno} Doppler
measurements \citet{Iess2018} solution.  However, in this solution the reported $J_n$
would include any tidal contribution ($C_{n0}$).  Thus for consistency, we subtract
the calculated Io $C_{n0}$ from the observed $J_n$ to obtain the target $J_n$. For
instance, the target $J_2=14696.51\times10^{-6}$ is used instead of the observed
$J_2=14696.57\times10^{-6}$. The interior models consist of an outer, molecular
envelope, and an inner, metallic envelope separated by a transition in which helium
is proposed to phase separate and rain out
\citep{stevenson-astropj-77-ii,Morales2013,Militzer2016}. Each model includes an
innermost spheroid representing a constant density central core with a fractional
radius $r/r_j=0.15$.  The outer and inner envelope are parameterized by a set of
parameters ($S_1$,$Y_1$,$Z_1$) and ($S_2$,$Y_2$,$Z_2)$ respectively. The helium rain
region  is treated as a smooth transition of each parameter between two pressures,
while the central core is treated separate from envelope parameters with its density
required to conserve the total mass of the planet. A subset of these models (A3--4)
also consider a dilute core with a constant enrichment of heavy elements ($Z_3$) with
a higher concentration than in the metallic envelope ($Z_2$). Since there is a
density tradeoff between parameters and a lack of constraints in the deep interior,
we assume $S_3=S_2$ and $Y_3=Y_2$ for simplicity.

In model A1 $J_2$ and $J_4$ are matched by iteratively tuning $S_1=S_2$ and $Z_2$,
and $J_6$ by tuning the pressure of helium rain onset.  In other models, the pressure
of helium rain onset is set to $P=95.4$ GPa, consistent with the \citet{Morales2013}
phase curve and  an outer envelope adiabat with $S=7.07$ $k_B$/electron based on
Galileo entry probe temperature measurements \citep{seiff-1998}.\footnote{ This
    pressure is not consistent with the envelope entropy for the 3 or 4 models
    presented here, but we treat $P=95.4$ GPa as a baseline, given multiple sources
of uncertainty for the onset of helium rain.  } In Model A2, $J_2$ and $J_4$ are
matched to the same procedure, while $J_6$ is  matched by tuning the jump in $\Delta
S=S_2-S_1$ across the helium rain layer. Finally, models A3 and A4 fit $J_6$ by
tuning the heavy element fraction of a dilute core, $Z_3$, with a fractional radius
of $r/r_j=0.7$ and $0.5$ respectively. Models A3 and A4 require dilute cores with
$Z_3$ = 0.12 and 0.30, respectively, compared to the heavy element fraction for the
deep envelope with $Z_2$ = 0.066 in A1.

As noted previously
\citep{Wahl2017a,Guillot2018}, such simple 3 or 4--layer models  require outer
envelopes hotter than expected based on the \textit{Galileo} entry probe
measurements \citep{seiff-1998}, when using the \citet{Militzer2013} equation of
state. These models are thus interpreted to give a rough, conservative estimate
of the range of tidal responses that might be expected from variability in an
interior density structure alone matching the low order zonal harmonics, rather than
seeking a single model to match all desired constraints.

\subsection{Influence of deep winds} \label{sec:winds}

The observed zonal harmonics are not due solely to a barotropic interior
profile, but will have contributions from deep winds
\citet{Kaspi2013,Kaspi2017b}. Due to the limitations of potential theory,
differential rotation can only be implemented fully consistently for cylinders
extending through the deep interior of the planet \citep{Wisdom2016}. It is
therefore, not possible to implement the more complex 3D wind profile expected
for Jupiter, in which the cylindrical flow velocities decay rapidly at depths
where conductivity becomes high enough for flows to couple to the planet's
magnetic field \citep{Cao2017}. We therefore, cannot self-consistently test
their effect on the tidal response \citep{Militzer2019a}. As a conservative
estimate, we consider interior models that fit to $J_2$-$J_6$ with the contributions from
various wind models omitted. The first class of wind models use the $\Delta J_n$
from \citet{Kaspi2018}, which used the thermal wind equation to optimize a decay
function for the observed surface wind profiles to match the odd zonal
harmonics. The second class of wind models represent a class of models that can
be found when the interior structure and wind profiles are optimized
simultaneously \citep{Militzer2019b}.

The second set of models in Tab. \ref{tab:models} are constructed identically to the
reference model (A1) but with the target $J_n$ chosen to be $J_{n,{\rm Juno}} -
\Delta J_n$, where $\Delta J_n$ is set to the contribution from the optimized deep
wind profiles of \citet{Kaspi2018}, for the longitude--independent (model B1) and
longitude independent (model B2) profiles. The wind profiles that
lead to these $\Delta J_n$ are incompatible with a potential theory based method like
CMS. In lieu of a consistent method for predicting the wind contribution to the tidal
response, we consider the case where the winds have no direct contribution to
$k_{nm}$ and only affect the model through modifying the target $J_n$ in the CMS
calculation, to be different from the reference model, A1. Since the true
contribution to the winds to $k_{nm}$ should partially offset this difference, this
treatment should lead to conservative estimate of the range values that might result
from models with winds.  In Section \ref{sec:galilean}, we demonstrate that this
range is indeed small compared to Juno's sensitivity.

Finally, we calculate the tidal response for a third set of interior models (C1-C4 in
Tab. \ref{tab:models}) selected from an ensemble of models generated by a
simultaneous optimization \citep{Militzer2019b} an interior structure and wind
profile using a Monte Carlo approach. These models relate wind velocity-depth
profiles to contributions in $J_n$ using the thermal wind equation (TWE) as presented
in \citet{Kaspi2018}. Whereas \citet{Kaspi2018} used a single reference interior
density structure found an optimized wind profile to match the odd zonal harmonics
($J_3$, $J_5$,\dots), models C1-C4 attempt to mach the even $J_n$ with a combined
contribution from the CMS interior structure and a wind profile using the TWE, while
also attempting to match the odd Jn with the same wind profile. The parameterization
of the interior structure in these models is similar to \citet{Wahl2017a}. In
practice these models are found able to match the \textit{Juno} $J_n$, while better
fulfilling the composition constraints from \textit{Galileo}, due to the added
freedom for the wind profile to account for a portion of the even $J_n$. These models
result in larger $\Delta J_n$ than \citet{Kaspi2018}, particularly for $J_6$, and
consequently lead to a slightly larger range in calculated $k_{nm}$.

\section{Results} \label{sec:results}

\subsection{Convergence behavior} \label{sec:convergence}

Here we study the convergence behavior of the CMS calculations for the
equilibrium tidal response. This is of particular relevance when considering the
higher degree Love numbers, as the tesseral moments of various degree, $n$, and
order, $m$, can exhibit very different convergence behavior. This must be
accounted for when reporting the numerical precision of a given $k_{nm}$
calculated using the CMS approach.

\begin{figure}[h!]
  \centering
    \includegraphics[width=0.5\textwidth]{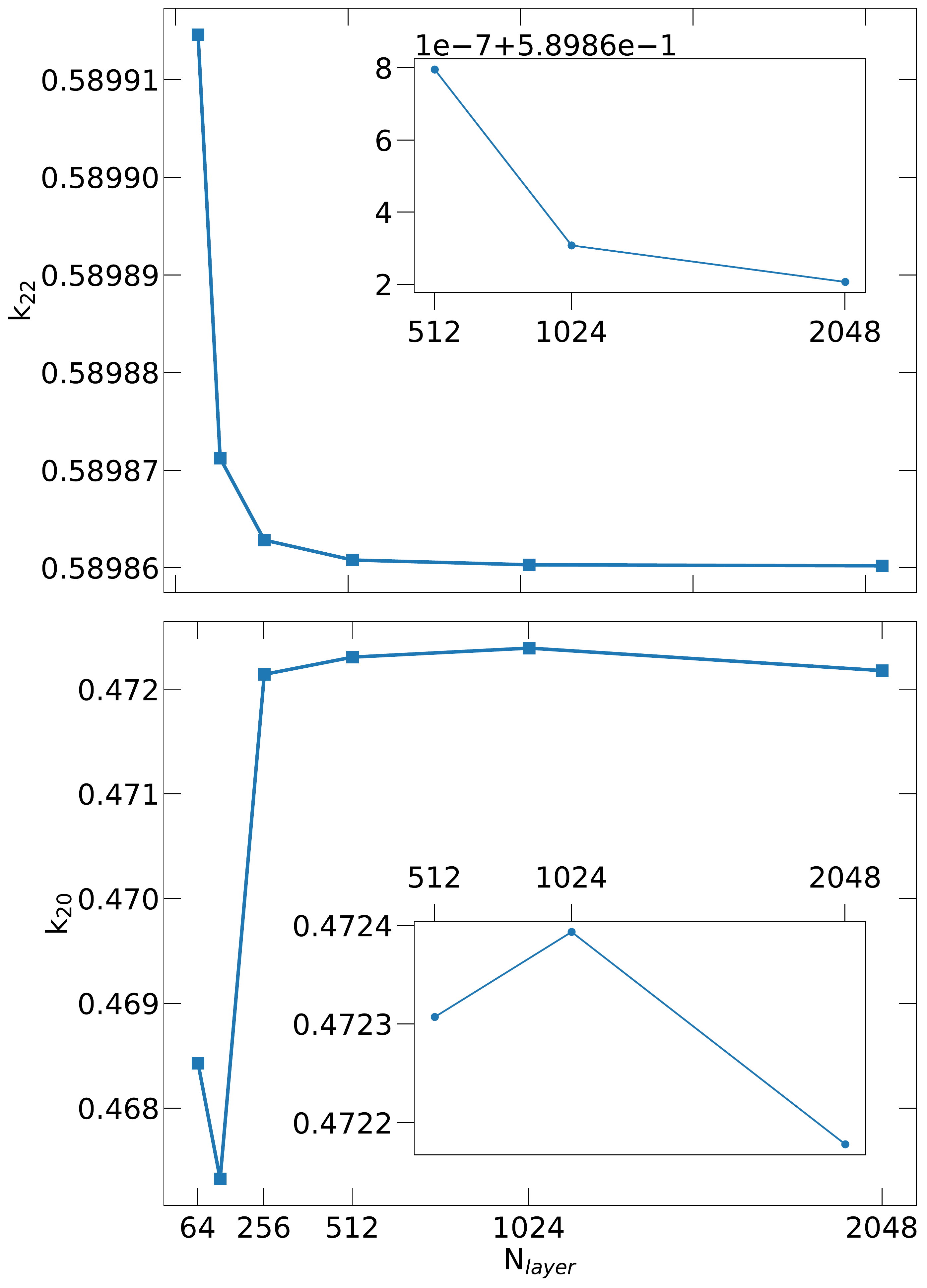}
\caption{ Convergence of Love numbers $k_{22}$ and $k_{20}$ with the number of
spheroids, $N_{L}$, in the CMS model. The numerical precision for Love numbers
of different degree and order are limited by their convergence at $N_{L}=512$.
The Love numbers of order $m=0$ are determined to significantly lower precision
that the other Love numbers.}
\label{fig:nl_converge}
\end{figure}

First, we study the discretization error by comparing the calculated $k_{nm}$
for models for an increasing number of spheroids, $N_l$ in the CMS calculation.
Fig. \ref{fig:nl_converge} shows the convergence of the calculated $k_{22}$ and
$k_{20}$  from 64 layers to the 512 layer model used in the rest of of the
paper, and then up to 2048 layers. The difference between 512 and 1024 layers in
the upper panel shows that 512 layers is sufficient to  derive $k_{22}$ to 6
significant digits. Meanwhile, $k_{20}$ is only converged to the level of
$\sim10^{-4}$. $k_{22}$ has a notably better convergence behavior that the
higher degree Love numbers, which converge to $\sim10^{-5}$ when $n=m$, but with
precision worsening as $n-m$ increases. Up to degree $n=5$, all $k_{nm}$ are
converged with $N_l$ to better than $2\times10^{-4}$. In the current
implementation, calculations more than 1000 layers are too computationally
expensive for  a detailed study, although the accelerated approach for the 2D
CMS approach \citep{Militzer2019a} could  be adapted to the 3D CMS method.

Next, we study the convergence behavior of the various $k_{nm}$ from a single
model through the iterative procedure. Whereas the Love numbers with order
$m\neq1$ are converged to at least the level of the discretization error after
25 iterations, those with $m=1$ require $\sim150$ iterations. This can be
attributed to the fact that the rearranging of the spheroid centers of mass
described in Sec. \ref{sec:update} is slower than the convergence of the shape
of the spheroid. In the original implementation of the CMS tidal response
\citep{Wahl2016,Wahl2017b}, the convergence of $k_{31}$ and $k_{51}$ were
essentially halted at a precision of $\sim10^{-2}$--$\sim10^{-3}$. This appears
to have a negligible effect on the results for the other $k_{nm}$ where
$n\neq1$. The convergence behavior over the course of the calculation is also
found to be nearly independent of the strength of the tidal perturber $q_{\rm
tid}$.

\begin{figure}[h!]
  \centering
    \includegraphics[width=0.6\textwidth]{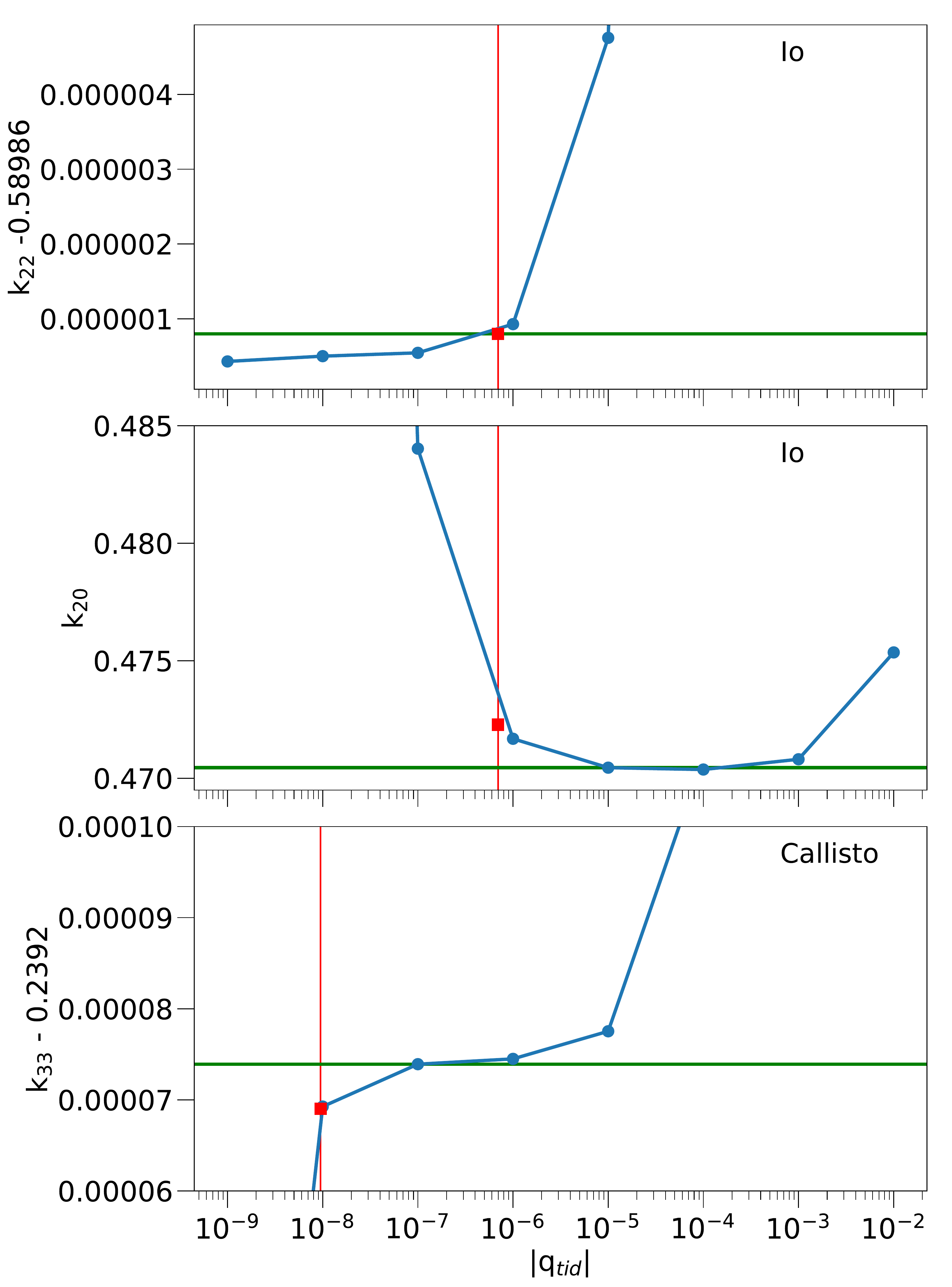}
\caption{ Representative examples of the convergence of various Love numbers as
a function of tidal perturber strength, $q_{\rm tid}$ (blue). The red vertical
line denotes the value of $q_{\rm tid}$ corresponding to the perturbing
satellite. Shown in red is the value obtained directly from the CMS simulation
with $|q_{\rm tid}|$ corresponding to the satellite. Shown in green is the
reported value for $k_{nm}$, which depends on whether the CMS simulation at the
correct $q_{\rm tid}$ resolves a value better than the estimate for the linear
regime. (Top) $k_{22}$ for a satellite at Io's orbital distance, which resolves
the non-linearity of the Love number. (Middle) $k_{20}$ for a satellite at Io's
orbital distance. In this case the best estimate of the linear regime occurs at
$q_{\rm tid,io}$. This is the case for all Love numbers of order 0 regardless
of satellite. (Bottom) For Callisto, the smaller magnitude of $q_{\rm tid}$
means the best estimate of $k_{33}$ is the estimate of the linear regime, even
though $k_{33}$ value is resolved directly for Io.  }
\label{fig:qtid_converge}
\end{figure}

Finally, we look at the convergence behavior of $k_{nm}$ with tidal perturber
strength, $q_{\rm tid}$.  Fig. \ref{fig:qtid_converge} shows three examples for
the convergence of a given  $k_{nm}$, where the satellite's orbital distance is
fixed, while $q_{\rm tid}$, or equivalently the satellite mass, is varied.
Starting from a magnitude of $q_{\rm tid}$ much greater than the corresponding
satellite value, the value of $k_{\rm nm}$ initially approaches a constant value
with decreasing $|q_{\rm tid}|$. This constant demonstrates the expected tidal
response in the limit of small perturbation. However, at the smallest values of
$|q_{\rm tid}|$, the value of $k_{nm}$ diverges again from the constant value, due
to the limited numerical precision of the calculation. In the case of Io's
$k_{22}$, the CMS calculation can clearly resolve a small non-linearity at the
corresponding  $q_{\rm tid}$.

However, the convergence behavior for other $k_{nm}$ with $q_{\rm tid}$ differ
significantly from $k_{22}$. From the very same CMS calculation at Io's orbital
distance, the value of $k_{20}$ at the satellite $q_{\rm tid}$ is affected by
this limited precision. In this case we cannot resolve any non-linearity in
$k_{20}$, and we instead report the linear-regime value, evaluated at the
$q_{\rm tid}\sim-10^{-5}$, where the change in $k_{20}$ between different $q_{\rm tid}$ is at a minimum. In
Fig. \ref{fig:qtid_converge} the reported $k_{nm}$ is shown in green. The
precision of this $k_{nm}$ is limited by the fact that it is derived by taking
the finite difference of Eq. \ref{eq:kn} between two calculations performed with
different $q_{\rm tid}$. This uncertainty in the calculated linear regime
$k_{nm}$  limits the precision of $k_{20}$ and the other $k_{n0}$, and is
included into the reported numerical uncertainty.

The precision of the higher degree $k_{nm}$ becomes increasingly limited as the
satellite becomes more distant due to the decrease in $q_{\rm tid}$. For example, the
reported $k_{33}$ for Callisto is the linear-regime result, whereas the larger
$q_{\rm tid}$ of Io or Europa allows $k_{33}$ to be calculated directly. These limits
to precision of the linear regime $k_{nm}$ calculation, are well below the potential
sensitivity of \textit{Juno}. The sensitivity of \textit{Juno} is discussed in detail
in Sections \ref{sec:anomaly} and \ref{sec:doppler}, but can be estimated taking a
relatively optimistic assumption that \textit{Juno} can detect a signal with a
maximum strength of $\sim1$ $\mu$gal. If the signal from $C_{20}$ considered in
Fig.\ref{fig:qtid_converge} were modified to account for a 1 $\mu$gal change in the
maximum anomaly, it would cause a corresponding  $\sim0.01$ change in $k_{20}$, which
is two orders of magnitude larger than the estimated numerical uncertainty. However,
directly using the value calculated at the satellite $q_{\rm tid}$ can result in a in
a significant miss-reporting of some $k_{nm}$, when the magnitude of $q_{\rm tid}$ is
sufficiently small.

\subsection{Tidal response from Galilean satellites} \label{sec:galilean}

Tab. \ref{tab:Io} tabulates the calculated Love numbers for the equilibrium
tidal response of Jupiter to Io. Independent calculations with the identical,
reference interior model (A1) were performed at the satellites perijove and
apojove, yielding a linear correction with orbital distance, $dk_{nm}/dR_{\rm
sat}$ in units of Jupiter's equatorial radius.

The Love numbers for a non-rotating analogue planet are also tabulated. As was
noted in previous work with the CMS method, $k_{nm}$ for a non-rotating planet
are independent of $m$, while for the rapidly rotating Jupiter these values
diverge as a result of rotational flattening.  The value of $k_{22}=0.5897$ for
the reference Jupiter model with rotation is significantly higher than the
non-rotating case with $k_{22}=k_{20}=0.5364$.

A similar disparity is seen when the CMS result is compared to the Radau-Darwin
approximation, commonly used for tides on bodies with less significant
rotational flattening. The normalized moment of inertia ($C/Ma^2$) can be
calculated directly from the CMS interior density structure
\citep{hubbardMilitzer2016}. The calculated moment of inertia $C/Ma^2=0.2639$,
would correspond to a $k{22}\sim0.524$ under this approximation.

As described in Section \ref{sec:convergence}, the table notes whether the CMS
calculation is able to resolve non-linear behavior with $q_{\rm tid}$. The numerical
uncertainty represents the estimated discretization error for the reported 512 layer
CMS calculation. In the cases where the non-linear behavior cannot be resolved at
Io's $q_{\rm tid}$, we report the linear regime result and include the corresponding
uncertainty in the reported numerical uncertainty. At the strength
of Io's perturbation all values of $k_{nm}$ with $m\neq 0$ can resolve non-linearity
up to degree $n=10$, while non-linearity is not resolved for any $k_{nm}$ with
$m=0$.

For Io specifically, we also considered the full suite of interior models (A1-C4)
tuned to fit the observed low order zonal harmonics $J_2$, $J_4$ and $J_6$, as
described in Sec. \ref{sec:interior}. These are tabulated as ``\textit{err
interior}'', and in most cases represent a range less than one order of magnitude
larger than the numerical precision.  Likewise, the range for each Love number from
the two sets of interior models constructed with consideration of the deep winds
(Sec. \ref{sec:winds}) are tabulated as ``\textit{err winds K18}'' and ``\textit{err
winds M19}''. Even with their density contributions completely omitted, the range of
$k_{nm}$ predicted for wind profiles differ from the reference model (A1) by an
amount smaller than could be observed by \textit{Juno}. The error
reported in ``$k_{nm}$ perijove'', includes the maximum deviations from the different
interior and wind profiles, along with the estimated numerical error and error
propagated from uncertainties in physical constants. The  For $k_{22}$, the
largest source of uncertainty comes not from interior structure or winds, but from
propagating the $\sim4$ km uncertainty on Jupiter's observed radius into $q_{\rm
rot}$, and thus to the rotational flatting of the body. In spite of our conservative
estimates of the various sources of uncertainty, the combined uncertainty remains
well bellow the expected sensitivity of \textit{Juno}. For instance, an optimistic
sensitivity of a $\sim1$ $\mu$gal, suggests and uncertainty two orders of magnitudes
larger than greatest disparity in $k_{22}$ between models with different interior
structures or winds. This suggests that the
$k_{nm}$ are extremely well determined by the low-order gravitational harmonics, and
thus provide little additional information to constrain the deep interior structure
of the planet.

Tab. \ref{tab:Io} also presents the value of $k_{nm}$ for the reference interior
model at apojove, along with the corresponding derivative with orbital distance,
$dk_{nm}/dR_{\rm sat}$. For $n<4$ the variation in $k_{nm}$ from orbital
distance is small compared to the numerical uncertainty and ranges from
different interior models.  However, in the case of $k_{42}$ the deviation of
$k_{nm}$ between apojove and perijove is much more substantial. The same is true
for other higher degree $k_{nm}$ where $m\neq n$, although in most cases this
effect is likely below detectability. The one possible exception is for
$k_{20}$, where, despite $dk_{nm}/dR_{\rm sat}$ being smaller than the reported
uncertainties, independent calculations yield a signal with consistent
amplitude, in which the observable $J_2$ varies by a magnitude similar to
\textit{Juno's} uncertainty \citep{Iess2018}.

For the reference interior model (A1), identical calculations were performed for
the three other Galilean satellites, and tabulated in Tabs.
\ref{tab:Europa}--\ref{tab:Callisto} and for the Sun in \ref{tab:Solar}. These
tables have been truncated to discount $k_{nm}$ that would produce signals well
below the detectability of \textit{Juno}; the full tables (up to $n=16$) can be
found in the supporting data. For these bodies, the reported uncertainty is the
numerical uncertainty plus the uncertainty on $q_{\rm rot}$, as reported for Io,
but without the additional ranges from different interior density profiles or
winds.

The qualitative behavior of $k_{nm}$ is similar between satellites, although the
precise values differ due to non-linearity in the response to both $q_{\rm tid}$ and
$R/r_{\rm J}$. As described in \citet{Wahl2016}, the rotational flattening of Jupiter
leads to a splitting of the $k_{nm}$ values from those calculated for a non-rotating
analogue planets. Most of the low-degree $k_{nm}$ agree with those presented in
\citet{Nettelmann2019}. Our calculated value for Io's $k_{22}$ matches theirs to
within our reported uncertainty. The largest disparities occur for $k_{nm}$ with
$m=1$, with their values for Io's $k_{31}$ and $k_{51}$ differing from those
presented here by $\sim30\%$ and $\sim15\%$ respectively. This disparity can likely
be attributed to the treatment of the average force and spheroid centers of mass
described in Sec. \ref{sec:update}, as they used the older CMS implementation that
suffered from the offset center of mass. The match between the independent
calculations also becomes poorer for $k_{nm}$ with higher degree $n$ and for more
distant satellites, which may reflect limitations of numerical precision with small
$|q_{\rm tid}|$ summarized in Sec. \ref{sec:convergence}. Nonetheless, the good
agreement supports our conclusion of the insensitivity of the equilibrium tidal
response to the details of the interior model, including those based on a different
equation of state.

The magnitude of the splitting becomes more significant with increasing degree
$n$ and orbital distance. It is also noteworthy that higher degree $k_{nm}$ with
$m\neq n$ show a much greater difference from the non-rotating analogue, than
the $k_{nm}$ with $m=n$. This can be related to the geometry of the tesseral
harmonics in Fig. \ref{fig:harmonics}, as harmonics with $m\neq n$ are those
that exhibit nodes in longitude and therefore map differently onto a flattened
spheroid, than those without such nodes.

\begin{figure}[h!]
  \centering
    \includegraphics[width=0.7\textwidth]{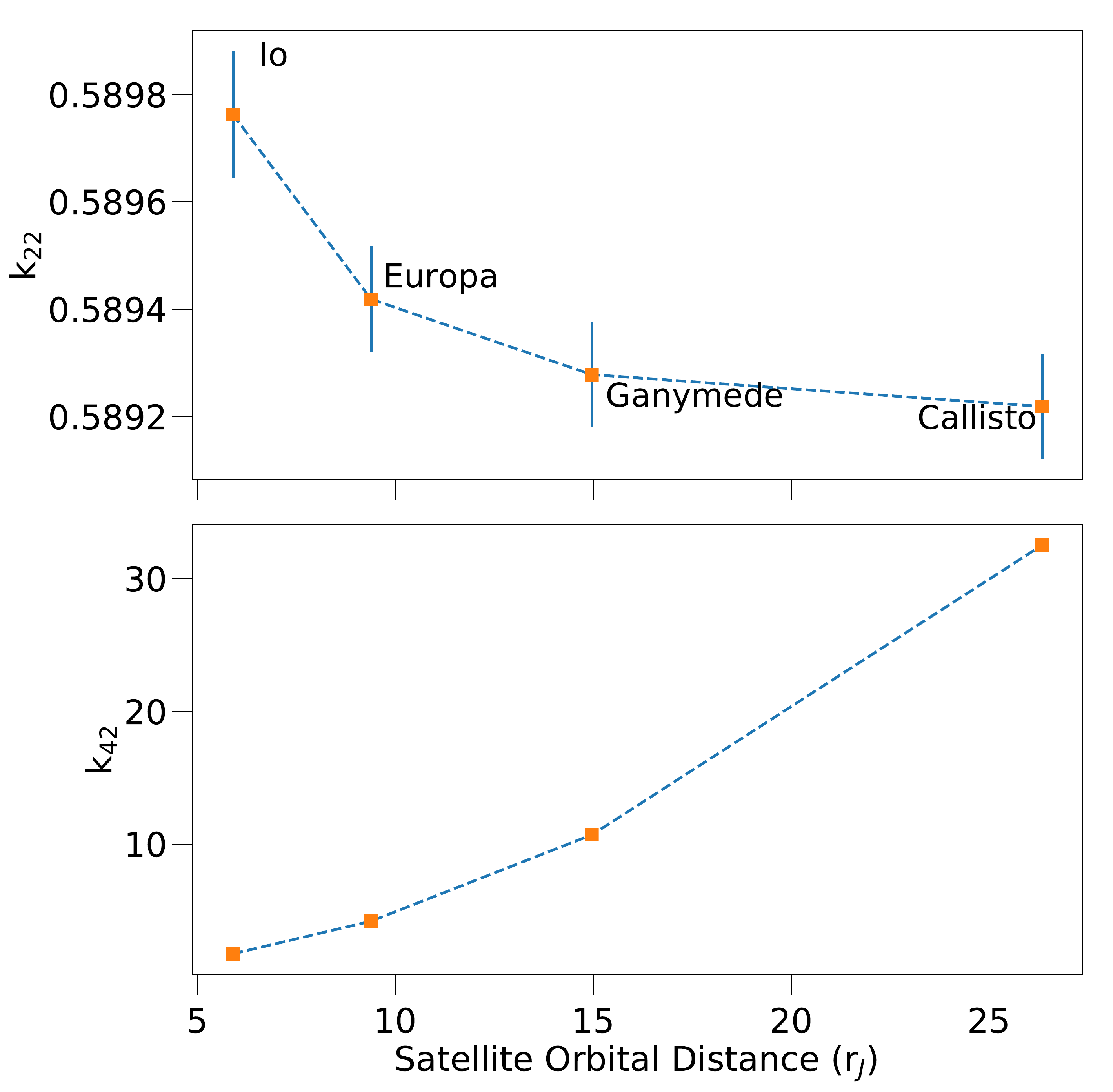}
\caption{ The dependence of Love numbers $k_{22}$ and $k_{20}$ on
the orbital distance of the four Galilean moons. The relative strength of the
effect becomes more significant for both larger degree $n$, and for larger $n-m$. The
effect is relatively small for $k_{22}$ (top panel) and more significant for $k_{42}$
(bottom panel). For Io, the errorbars show the range of $k_{nm}$ for interior models
matching the observed $J_{n}$.} \label{fig:knm_distance}
\end{figure}

Fig. \ref{fig:knm_distance} shows an example of how two such $k_{nm}$ vary with
orbital distance. For $k_{22}$ the change is relatively modest, with the value
predicted for Europa differing by only $\sim 3\times10^{-4}$, likely too small
of a difference to be observable to Juno, but still an order of magnitude larger
than the uncertainty introduced from considering different interior models that
fit the observed $J2$-$J6$. In contrast, $k_{42}$ at Callisto's orbital distance
is over 18 times larger than for Io. As a result, the satellite dependent
equilibrium tidal response is most readily observable for $k_{42}$, even though
the magnitude of the signal corresponding to $k_{22}$ and $k_{33}$ are larger.
In the case of the tide raised by the Sun, Tab. \ref{tab:Solar}, the splitting
of the calculated higher--degree $k_{nm}$ become quite extreme, but the
corresponding harmonic $C_{nm}$ strength decays rapidly with degree $n$, such
that all $C_{nm}$ with $n>2$  are far below levels detectable by \textit{Juno}.
The lack of a substantial satellite dependence for $k_{22}$ does not rule out
the possibility of a satellite dependent dynamic contribution
\citep{Notaro2019}. In fact, given the small difference in equilibrium $k_{22}$
calculated for the various satellites, measurement of disparate $k_{22}$ at the
different satellite orbital frequencies might be taken as evidence for a dynamic
tidal contribution.

In order to model the complete equilibrium tidal bulge on Jupiter, the
contributions from all four of the satellites must be taken into account. Since
for each satellite $q_{\rm tid} \ll q_{\rm rot}$, it is a reasonable assumption
to treat the full tesseral harmonic as a linear superposition of contributions
from the separate satellites.  Rearranging Eqs. \ref{eq:qtid} and \ref{eq:kn},
the contribution of a satellite to the tesseral harmonics is given by
\begin{eqnarray} \label{eq:phase}
    C_{nm} = C_{nm,0} \cos \left( m ( \Phi - \Phi_0 ) \right) \\
    S_{nm} = C_{nm,0} \sin \left( m ( \Phi - \Phi_0 ) \right),
\end{eqnarray}
where $\Phi$ and $\Phi_0$ are the phase of the satellite and a reference phase, and
\begin{equation} \label{eq:cmm_from_knm}
    C_{nm,0} = 2\frac{(n-m)!}{(n+m)!} \left( \frac{m_{\rm s}}{M} \right)
    \left( \frac{r_{\rm eq}}{R} \right)^{n+1} P_n^m(0) k_{nm},
\end{equation}
with $k_{nm}$ at a given  orbital separation $R$ given by the various
tables. $R$ varies over the course of an elliptical orbit, leading to small
changes in both $r_{\rm eq}/R$ and $q_{\rm tid}$.

\begin{figure}[h!]
  \centering
    \includegraphics[width=0.7\textwidth]{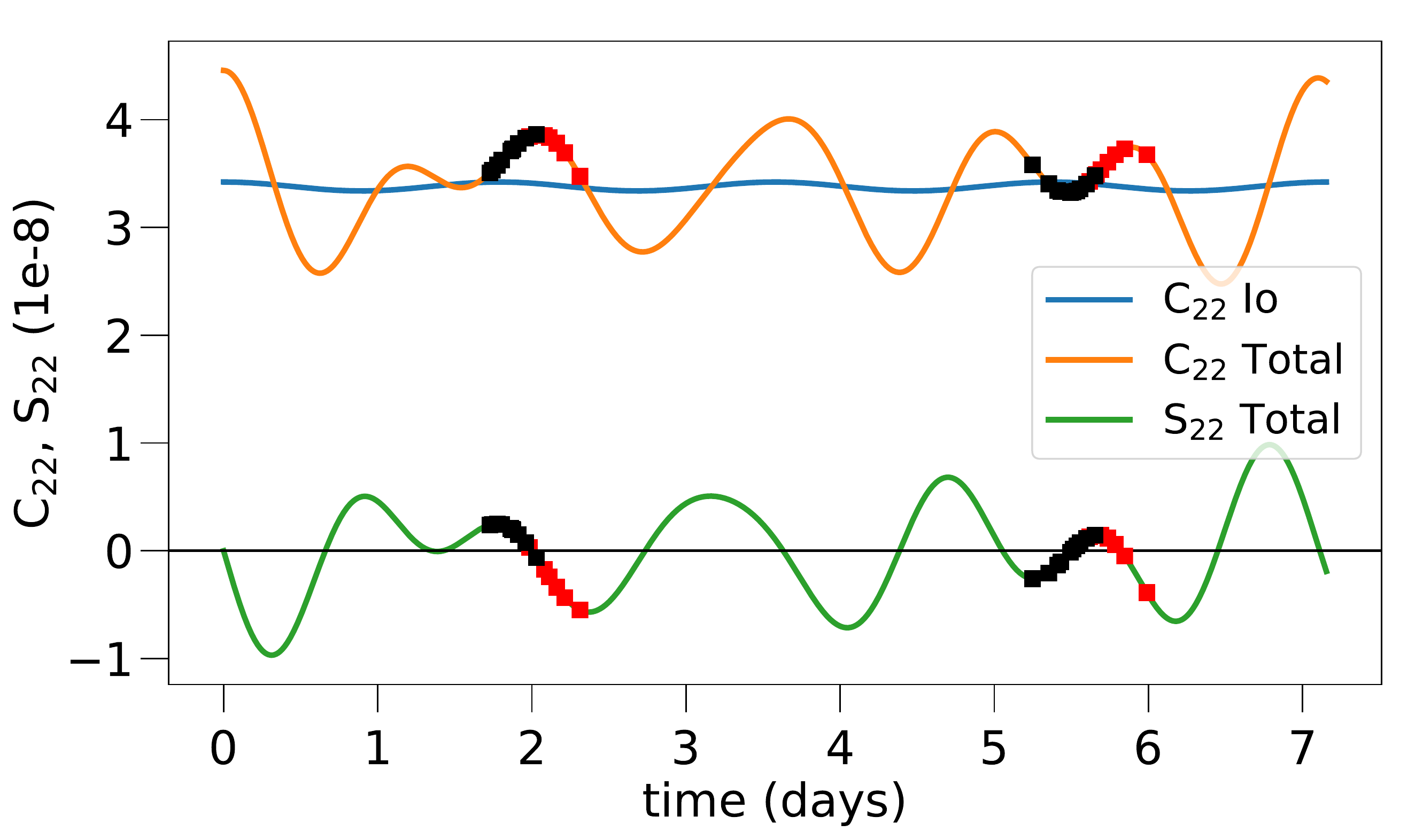}
\caption{Combined contribution to the tesseral gravity harmonics $C_{22}$
(yellow) and $S_{22}$ (green) from all Galilean moons over the course of one
orbit of Ganymede (4 orbits of Io), with $t=0$ at the inferior conjunction of Io
and Ganymede. The coordinates are chosen such that Io contributes only to
$C_{nm}$, with $S_{nm}$ contributions arising from the other satellites. The
contribution from $C_{nm}$ from Io alone is shown in blue. The squares show the
point in the cycle at \textit{Juno} perijove, with completed PJ1-PJ21 in black
and projected PJ22-PJ35 in red.}
\label{fig:c22_s22}
\end{figure}

\begin{figure}[h!]
  \centering
    \includegraphics[width=0.7\textwidth]{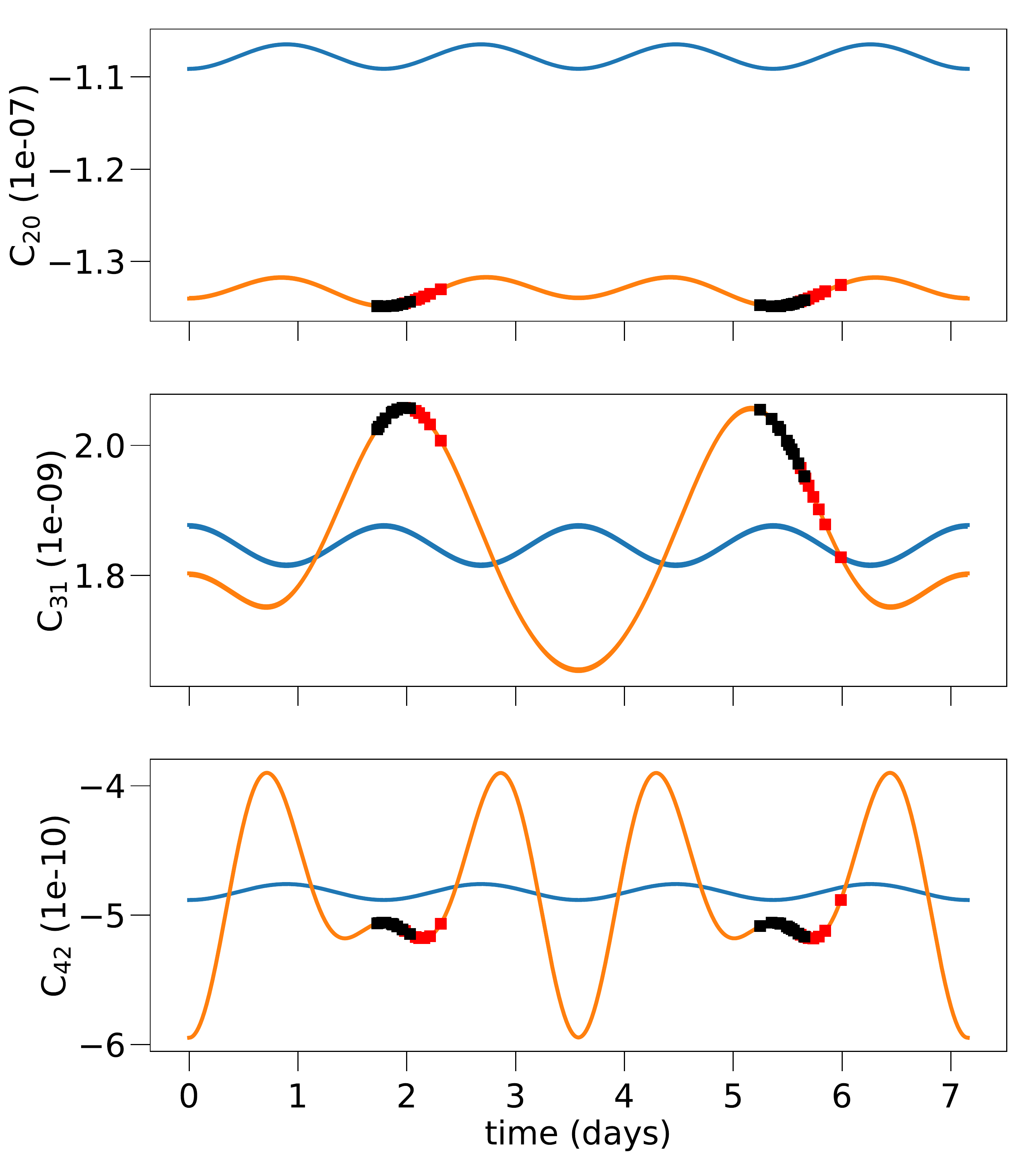}
\caption{Combined contribution to $C_{nm}$, (yellow) from all Galilean moons
over the course of one orbit of Ganymede, as in Fig. \ref{fig:c22_s22}. The
contribution from $C_{nm}$ from Io is shown in blue. The squares show the point
in the cycle at \textit{Juno} perijove, with completed PJ1-PJ21 in black and
projected PJ22-PJ35 in red. } \label{fig:cnm_total}
\end{figure}

Since the three satellites with the strongest tidal perturbations are in a 4:2:1
orbit resonance, the time-dependent, equilibrium tidal bulge is dominated by a
signal that repeats once every orbit of Ganymede. It is convenient for
visualization to set the reference phase $\Phi_0$  in Eq. \ref{eq:phase} to that
of Io's orbit. In this frame of reference the primary contribution to the tidal
bulge from Io remains fixed, contributing only to $C_{nm}$, while contributions
from the other satellites cause temporal variations in $C_{nm}$ and $S_{nm}$
from that baseline value. Fig. \ref{fig:c22_s22} shows this repeating pattern in
the tidal response in terms of $C_{22}$ and $S_{22}$, over a single orbit of
Ganymede with $t=0$ taken to be at inferior conjunction of Io and Ganymede.
Fig. \ref{fig:cnm_total} shows the corresponding pattern for $C_{nm}$ of other
selected low--degree harmonics.

Due to the coincidental match of \textit{Juno}'s orbital period with this
resonance, the spacecraft perijoves occur within a limited range of $\Phi_{\rm
Io}-\Phi_{\rm Europa}$. The point in the orbital cycle for each \textit{Juno}
perijove are shown on the $C_{nm}$ curves in Figs. \ref{fig:c22_s22} and
\ref{fig:cnm_total}, for both the completed (PJ1-PJ21) and currently projected
(PJ22-PJ35) perijoves. If the \textit{Juno} spacecraft orbit is altered during
the extended mission to allow for a larger range of $\Delta \Phi$, then the
sensitivity of the \textit{Juno} gravity solution to the satellite  specific
$k_{nm}$ would increase.

While $k_{20}$ is not directly observable through the means that $k_{nm}$ with
$m>0$ are, the top panel of Fig. \ref{fig:cnm_total} suggests an indirect means
of measuring it. The  equilibrium $C_{20}$ is varies with Io's orbital
distance. The total equilibrium tidal contribution to $J_2$ is
$\sim6.6\times10^{-8}$ and would be embedded in much larger contributions from
the interior or winds. However the variations from Io's orbital eccentricity
would cause the total $J_2$ to vary by $\sim1.5\times10^{-9}$. This is roughly
an order of magnitude smaller than the reported $J_2$ uncertainty of the
\citet{Iess2018} gravity solution, but may become detectable as the uncertainty
decreases over the course of the mission.

\subsection{Gravity anomaly from the Love numbers} \label{sec:anomaly}

\begin{figure}[h!]
  \centering
    \includegraphics[width=0.7\textwidth]{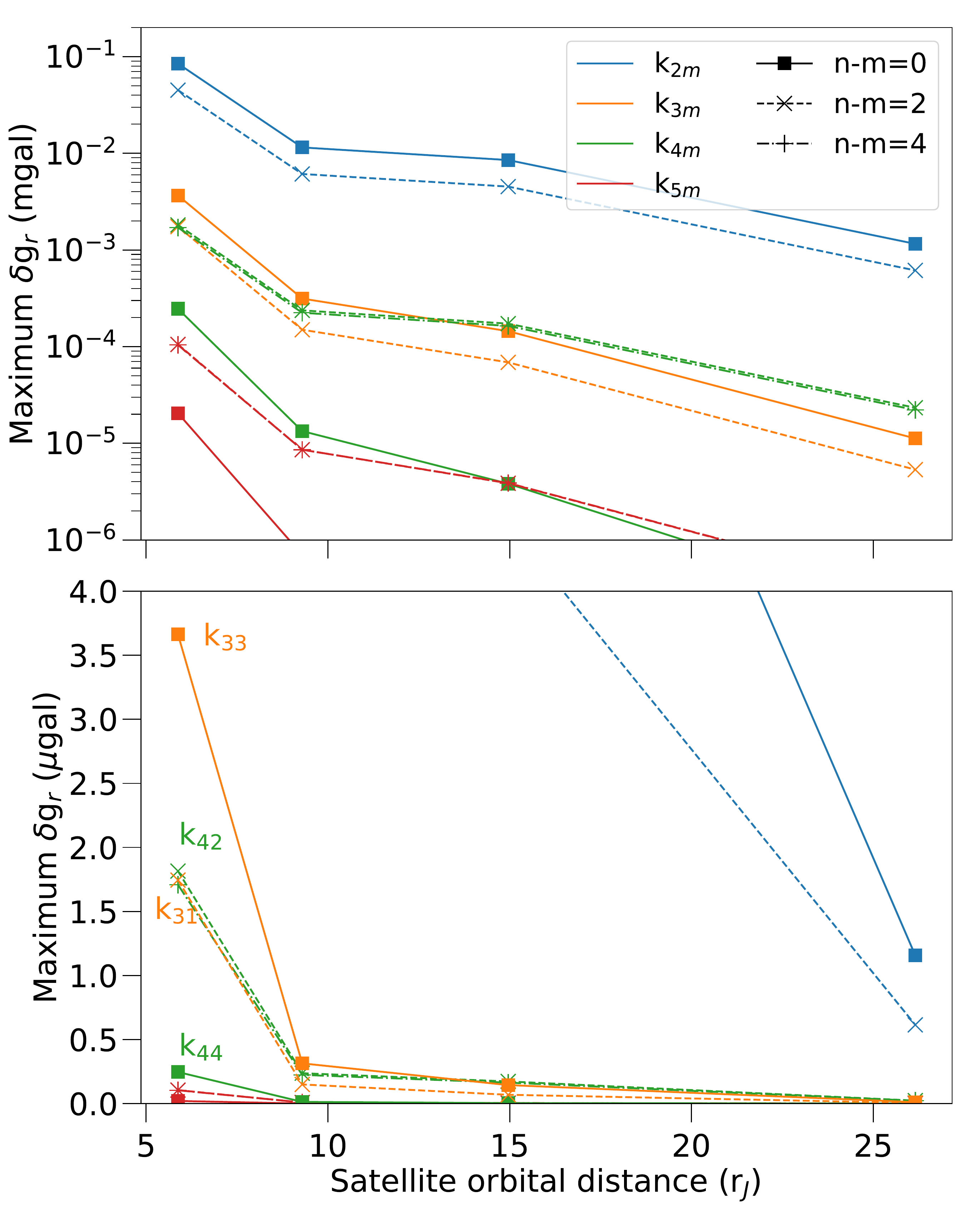}
\caption{ Top panel: maximum gravity anomaly, $\delta g_r$, at the surface
resulting from the calculated $k_{nm}$ for each Galilean satellites. Bottom
panel: same but on a linear scale at the \textmu gal level. The most readily
observable $k_{nm}$ with $n>2$ are labeled.
 } \label{fig:anomaly}
\end{figure}

A straightforward way of estimating the relative detectability of a given
$k_{nm}$ is to calculate the maximum gravitational anomaly associated with that
specific Love number. Starting with the corresponding tesseral gravity harmonic
$C_{nm}$ from Eqn. \ref{eq:cmm_from_knm}, the gravity anomaly at the
sub-satellite point ($\mu=0$, $\phi=0$) that results from adding that tesseral
signal to an axisymmetric gravity solution is
\begin{equation}
\delta g_r = C_{nm} \frac{P^m_n(0)}{n+1}\frac{G M_J}{r_J^2}.
\end{equation} \label{eq:anomaly}
Fig. \ref{fig:anomaly} shows the relative magnitude of the anomaly for various
$k_{nm}$ calculated for each of the four satellites at their perijove. Io's
$k_{22}$  yields a $\delta g_r \sim 0.085$ mgal, while Europa and Ganymede's
$k_{22}$ have a corresponding  $\delta g_r$ roughly one order of magnitude
smaller, and Callisto an order of magnitude smaller yet. In the lower panel of
Fig. \ref{fig:anomaly}, we see that the most readily detectable higher order
Love numbers, $k_{33}$, $k_{42}$ and $k_{31}$, have $\delta g_r$ values on the
order of a $\mu$gal.

We can attempt to predict which Love numbers \textit{Juno} is sensitive to by
comparing the calculated gravity anomaly magnitude with an estimate for
observational uncertainty. For instance, Fig. 3 of \citet{Iess2018} presents the
residual gravitational accelerations for a single close approach of the
spacecraft, during which the minimum uncertainty is on the order of $\sim0.1$
mgal. Using this value in conjunction with Fig. \ref{fig:anomaly}, we would thus
predict \textit{Juno} to be sensitive to only $k_{22}$ and $k_{20}$ for Io, and
not to any $k_{nm}$ from the other satellites. In the following section we
consider instead the sensitivity to a time-integrated signal from the calculated
equilibrium tides.

\subsection{Sensitivity of \textit{Juno} Doppler measurements to the calculated Love numbers} \label{sec:doppler}

The \textit{Juno} Gravity Science experiment uses measurements of the Doppler
shift of the radio signal transmitted by the spacecraft to determine the time
history of the spacecraft velocity projected onto the direction to the Earth
tracking station, $\dot\rho$. The velocity component is measured with accuracy
of about $\sigma_{\dot\rho} = 5 \mu m s^{-1}$ for averaging times of one minute,
limited primarily by fluctuations in the water content in the Earth's
troposphere \citep{Asmar2017}. The velocity measurements are used to estimate
corrections to models of the forces acting on the spacecraft, including
Jupiter's equilibrium gravity field and its tidal perturbations characterized by
the Love numbers. In order to estimate Love numbers $k_{nm}$ from the Doppler
measurements, we calculate the partial derivative for the change in velocity per
unit change in the value of the Love number, $\partial\dot\rho/\partial k_{nm}$
for each measurement time. Given a calculated value of $k_{nm}$, the magnitude
of the predicted velocity change due to each Love number is
\begin{equation}
\dot\rho_{pred} = k_{nm}\frac{\partial\dot\rho}{\partial k_{nm}}.
\end{equation}

\begin{figure}[h!]
  \centering
    \includegraphics[width=0.7\textwidth]{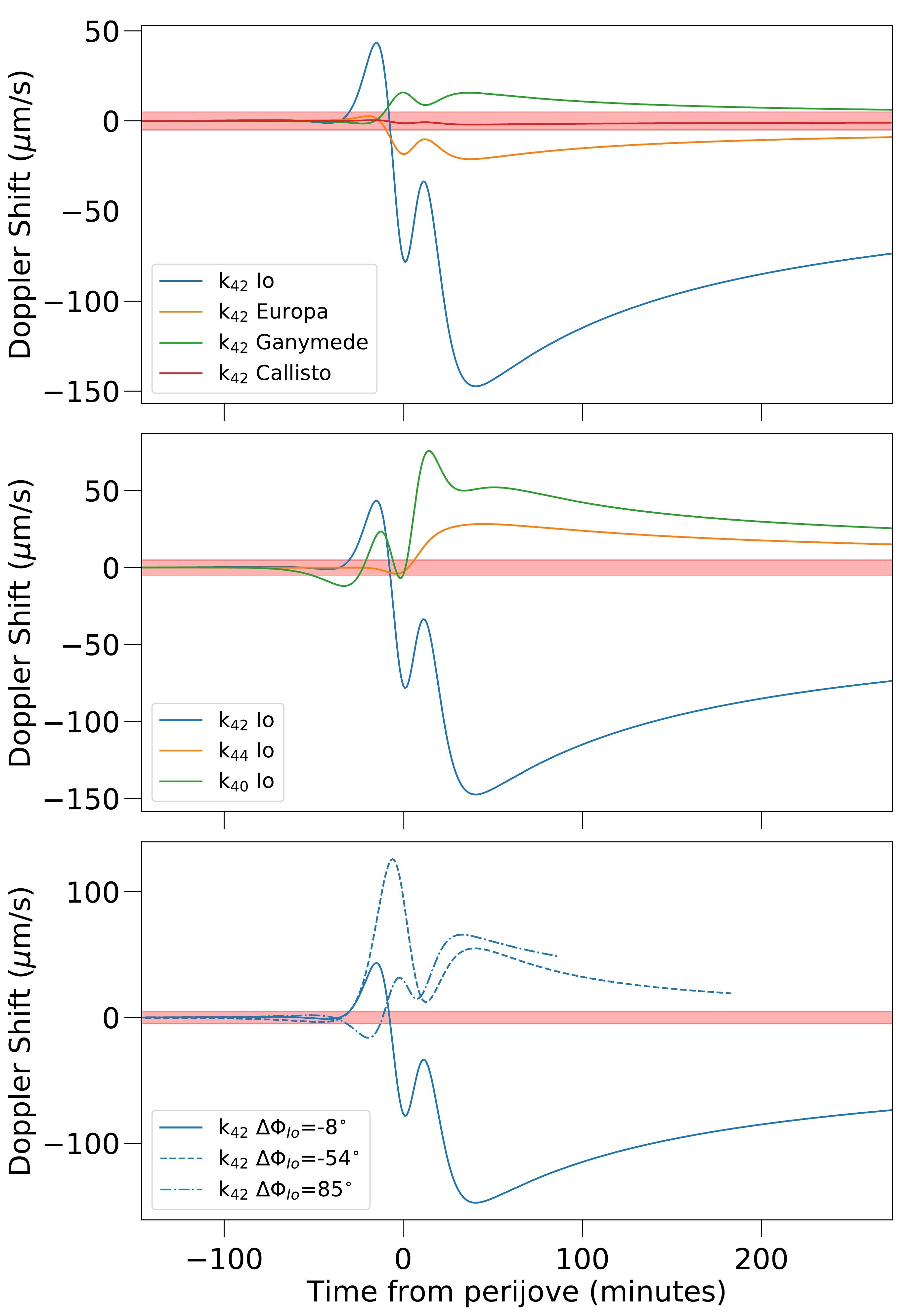}
\caption{ Top panel: Doppler shift as a function of time near \textit{Juno}
perijove resulting from adding a equilibrium $k_{42}$ with the magnitude for each
Galilean satellite from this paper at their PJ13 positions. The red filled
region shows the noise level for data about the optimized gravity solution for this
perijove. Middle panel: Same as top panel but showing the $k_{44}$, $k_{42}$ and
$k_{40}$ for Io. Bottom Panel: same as top panel, but comparing $k_{42}$ from Io
for PJ13, PJ06 and PJ03, during which the difference in longitude between
\textit{Juno} and Io ($\Delta \Phi_{\rm Io}$) is -8$^\circ$, -54$^\circ$, and
85$^\circ$, respectively. } \label{fig:doppler_residual}
\end{figure}

The top panel of Fig. \ref{fig:doppler_residual} shows the predicted Doppler
velocity, $\dot\rho_{pred}$, as a function of time for the thirteenth closest
approach of the \textit{Juno} spacecraft to Jupiter (labeled PJ13), on 24 May
2018, due to the tides raised on Jupiter by the four Galilean satellites, each
characterized by the calculated values of the Love number $k_{42}$. The signals
are well above the measurement noise level, $\sigma_{\dot\rho}$ , for about one
hour centered on closest approach and, as expected, Io raises the strongest
tidal effect on Jupiter. Equivalent signatures from $k_{22}$ are nearly 200
times larger than those from $k_{42}$, therefore we choose not to display both
effects, as they are difficult to display using the same scale for the signal
strength. Eventually the values of the Love numbers will be adjusted to best fit
the measurements from several \textit{Juno} perijoves. While the data are still
being calibrated, we use the theoretical values of the Love numbers, while
estimating the larger equilibrium gravitational signature to reduce the chance
of mismodeling of the tide signatures corrupting the equilibrium gravity field
estimate.

The middle panel of Fig. \ref{fig:doppler_residual} shows the degree-4 tidal
signal from Io during PJ13, for different values of the order $m$. Likewise, the
signal of these parameters are well above the noise cut-off level, hence we
expect them to affect the reduction of the \textit{Juno} Doppler data and
require them to be properly modeled. Furthermore, the predicted Doppler signal
from tides is a function not only of the Love number and tide raising body, but
also of the \textit{Juno}-Io phase angle (bottom panel).

\begin{figure}[h!]
  \centering
    \includegraphics[width=0.7\textwidth]{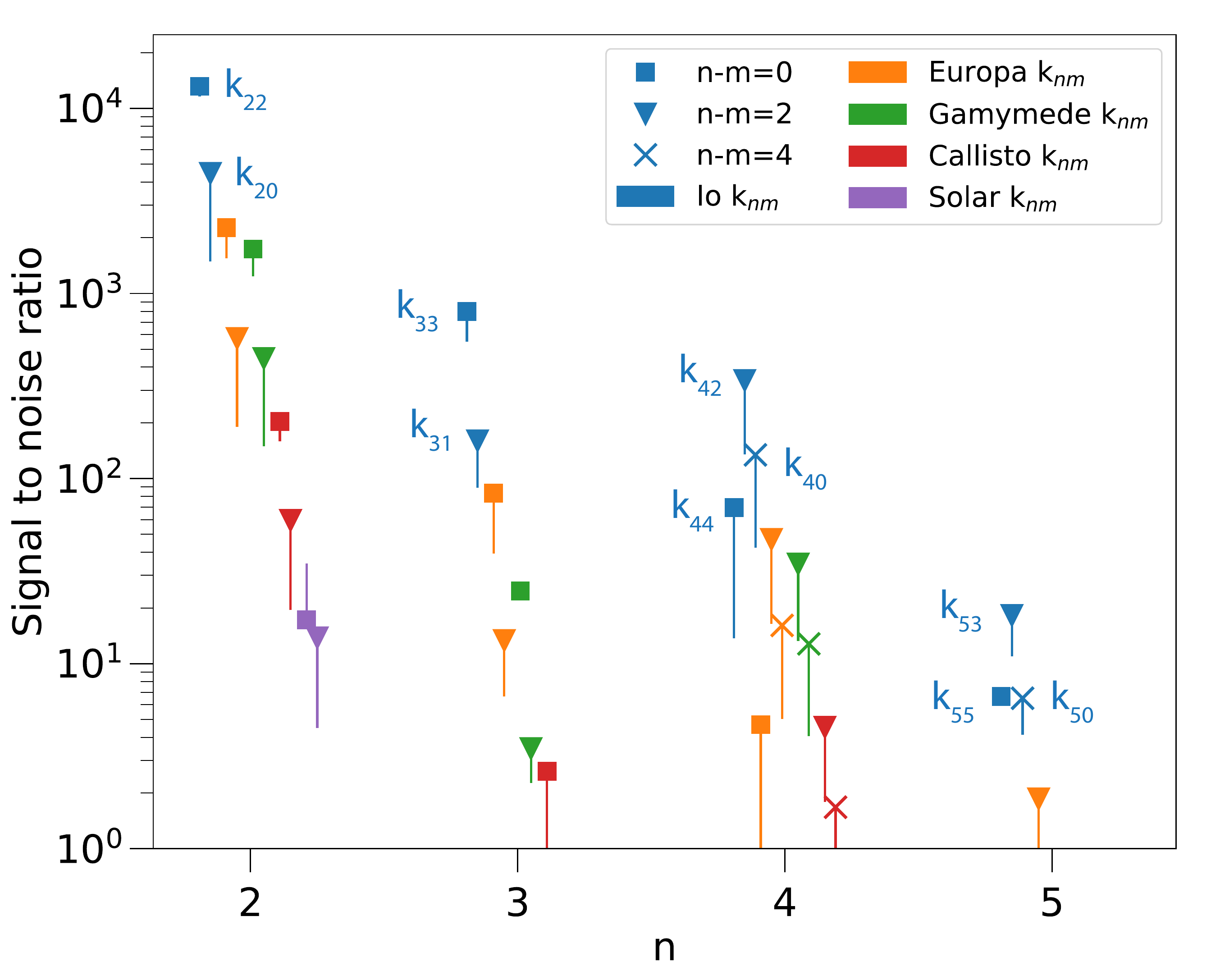}

\caption{Signal to noise ratio for time-integrated Doppler shift profiles (Fig.
\ref{fig:doppler_residual}) resulting from adding the calculated equilibrium
$k_{nm}$ for each Galilean satellite and the Sun from this paper at their PJ13
positions. During PJ13 \textit{Juno} was close to the Io sub-satellite point
($\Delta \Phi_{\rm Io}=-8^\circ$). The error bars show the range of $SNR$
including two other close approaches PJ06 and PJ03, with $\Delta \Phi_{\rm
Io}=-54^\circ$ and  85$^\circ$, respectively.
    } \label{fig:doppler_snr}
\end{figure}

In order to compare the size of the effects of the different Love numbers, Fig.
\ref{fig:doppler_snr} shows the signal-to-noise ratio (\textit{SNR}) during
\textit{Juno} thirteenth perijove (PJ13) for Love numbers $k_{nm}$. The
\textit{SNR} is defined here as the ratio of the nominal value of the Love
number and the uncertainty in its estimated value, $\sigma_{k_{nm}}$, from the
Doppler data when only the Love number is estimated. The uncertainty is given by
\begin{equation}
\sigma_{k_{nm}} = \frac{\sigma_{\dot\rho} }{\sqrt{\sum(\partial\dot\rho/\partial
k_{nm})}},
\end{equation}%
where the summation is over all Doppler measurements for a single perijove. The
\textit{SNR} is then given by
\begin{equation}
\mathit{SNR}_{nm} = \frac{k_{nm}}{\sigma_{k_{nm}}}.
\end{equation}

This definition of the \textit{SNR} allows a comparison of the amplitude of the
effect of the different equilibrium Love numbers on the Doppler measurements.  Fig.
\ref{fig:doppler_snr} shows that the SNR for Love numbers of degree and order 3
through degree and order 5 ranges from 2 to 2400, hence most of them are, in
principle, detectable in the Juno data. In general, the Love numbers cannot be
independently estimated because their signatures in the Doppler data are not
orthogonal to the other Love numbers, and are not orthogonal to the equilibrium
gravity signature or several other parameters describing other force
models. Therefore, \textit{Juno} is likely to be able to accurately estimate
values and uncertainties for Love number with \textit{SNR} greater than
$\sim$100, but inclusion of nominal values for lower \textit{SNR} Love numbers
is important to avoid aliasing of small tidal signals into other gravity
parameters.

Using this conservative threshold for \textit{SNR}, Fig. \ref{fig:doppler_snr}
suggests that \textit{Juno} will be able to estimate Io's $k_{33}$, $k_{31}$,
and $k_{42}$, with $k_{44}$ lying near the threshold. Likewise, it suggests that
$Juno$ may be able to estimate $k_{22}$ for the other three Galilean satellites.
The solar tide is below the threshold with its $k_{22}$ having a $\mathit{SNR}
\sim 30$, and the next most significant tidal perturber, Jupiter's satellite
Amalthea, yields an $SNR$ an order of magnitude smaller than the Sun. If the
signals could be sufficiently separated down to an $\mathit{SNR} \sim 10$, then
degree 3 and 4 Love numbers might be detected for Europa and Ganymede, along
with $k_{53}$ for Io. The conclusions of \textit{SNR} analysis are, therefore,
more optimistic towards the number of detectable $k_{nm}$ than would be
predicted from the gravity anomaly magnitudes alone (Sec.
\ref{sec:anomaly}). Comparing these predictions to the lower panel of Fig.
\ref{fig:anomaly}, suggests that \textit{Juno} is sensitive to tides with a
maximum gravity anomaly as low as $\sim 1 \mu gal$.

\section{Conclusions} \label{sec:Conclusions}

In this work, we calculate the equilibrium tidal response of Jupiter to its four
Galilean moons and the Sun. We present these as a series of tables  that report
the $k_{nm}$ for each body, characterizing their dependence on orbital distance,
and the estimated uncertainty from the numerical method, physical parameters,
interior  density structure and winds. We find an equilibrium $k_{22}=0.58976
\pm 0.0001$ for Io, consistent with previous calculations
\citep{Wahl2016,Nettelmann2019}, that is remarkably insensitive to details of
the interior structure model, once fitted to the low degree axisymmetric gravity
solution ($J_2$, $J_4$, and $J_6$) from \textit{Juno} \citep{Iess2018}. This
means that measurement of $k_{22}$ by \textit{Juno} will not yield additional
constraints on the interior structure via the equilibrium tidal response.
However, this insensitivity to interior models also means that \textit{Juno} has
the opportunity to unambiguously detect dynamic contributions to the tidal
response. Should dynamic contributions be detected, they may yield independent
information regarding interior structure or processes, although a comprehensive
theoretical predictions for such a dynamic response have not been performed.

We introduce improvements to the CMS method for tidal response calculations, which eliminate the
previously described center of mass shift resulting from the original
implementation. The improvement allows us to correctly resolve the Love numbers
of order $m=1$ (i.e. $k_{31}$ and $k_{51}$). We find that for the tides
experienced by Jupiter, that the predictions of the other $k_{nm}$ using the
previous method \citep{Wahl2016,Wahl2017b,Nettelmann2019} are consistent, at
least within the expected sensitivity of \textit{Juno}. It remains to be shown
whether the improved method has a more profound affect on predictions for
close-in extrasolar planets, where the tidal perturbations can be several
orders of magnitude stronger.

In Sec. \ref{sec:doppler} we studied the sensitivity of the \textit{Juno}
Doppler measurements to the calculated equilibrium tidal response. By finding
the signal to noise ratio for the calculated $k_{nm}$, we show that
\textit{Juno} is sensitive to both the high-degree ($n>2$) Love numbers of Io,
as well the $k_{2m}$ of Europa, Ganymede, Callisto and the Sun. This is
important for two reasons: First, it motivates the need for inclusion of the
higher order tidal components in the analysis and interpretation of
\textit{Juno} Doppler data. The signals from these higher order tides are
sufficiently large, that mischaracterizing them could lead to their
misinterpretation as contributions from from another source (i.e. interior
density structure or deep winds). Second, they suggest that multiple $k_{nm}$
may be detectable by \textit{Juno}. In principle, this could provide a test of
the theoretical predictions for the rotationally induced splitting and orbital
dependence of the equilibrium $k_{nm}$ \citep{Wahl2017b}. They may also offer
independent measurements from $k_{22}$ to detect or characterize a dynamic
contribution to the tides.

\section*{Acknowledgments}
The work of Sean Wahl and Burkhard Militzer was carried out at the University of
California, Berkeley, with the support of the National Aeronautics and Space
Administration, Juno Program. The work of Marzia Parisi and William M Folkner was
carried out at the Jet Propulsion Laboratory, California Institute of Technology,
under a contract with the National Aeronautics and Space Administration. Government
sponsorship acknowledged.






\bibliographystyle{aasjournal}                       
\bibliography{jupiter}



\floattable
\begin{deluxetable}{c|crl|crl}
\tabletypesize{\scriptsize}
\tablecolumns{7}
\tablewidth{0.5\textwidth}
\tablecaption{Jupiter Tidal Parameters}
\tablehead{
    Body & Parameter &  & & Parameter & & }
\startdata
Jupiter & $GM$ & 126686534.911 & km$^3$/s$^2$ & $q_{\rm rot}$ & $0.089195
\pm 1.5\times10^{-5}$ & \\
                  & $r_{eq}$ & $71492. \pm 4.$ & km & & &\\
                  & $T_{\rm rot}$ & 0:9:55:29.711 & D:H:m:s & & &\\
\tableline
Io & $GM$ & 5962.0 & km$^3$/s$^2$ & $R_{\rm pj}$ & 5.87534  & r$_J$ \\
              & $a$ & 421769. & km        & $R_{\rm aj}$ & 5.92372 & r$_J$ \\
              & $e$ & 0.0041 & & $q_{\rm tid,pj}$ & -6.87587$\times10^{-7}$ &  \\
              & & & & $q_{\rm tid,aj}$ & -6.79199$\times10^{-7}$ &  \\
\tableline
Europa & $GM$ & 3201.6  & km$^3$/s$^2$ & $R_{\rm pj}$ & 9.29196 & r$_J$ \\ 
                 & $a$  & 671079. & km           & $R_{\rm aj}$ & 9.48158 & r$_J$ \\
                 & $e$ & 0.0101  &                     & $q_{\rm tid,pj}$ & -9.44995$\times10^{-8}$ &  \\
                 &     &   &                      & $q_{\rm tid,aj}$ & -9.11439$\times10^{-8}$ &  \\
\tableline
Ganymede & $GM$ & 9891.0 & km$^3$/s$^2$ & $R_{\rm pj}$ & 14.95833 & r$_J$ \\
              & $a$ & 1070042.8 & km              & $R_{\rm aj}$ & 14.97629 & r$_J$ \\
              & $e$ & 0.0006  &                         & $q_{\rm tid,pj}$ & -6.99812$\times10^{-8}$ &  \\
              &     &   &                         & $q_{\rm tid,aj}$ & -6.97297$\times10^{-8}$ &  \\
\tableline
Callisto & $GM$ & 7181.3 & km$^3$/s$^2$  & $R_{\rm pj}$ & 26.15424 & r$_J$ \\
              & $a$ & 1883000. & km                & $R_{\rm aj}$ & 26.52298 & r$_J$ \\
              & $e$ &  0.007 &                          & $q_{\rm tid,pj}$ & -9.50535$\times10^{-9}$ &  \\
              &     &   &                          & $q_{\rm tid,aj}$ & -9.11439$\times10^{-9}$ &  \\
 \tableline
Sun & $GM$ & 132712440041.93938 & km$^3$/s$^2$ & $R_{\rm pj}$ & 10357.8 & r$_J$ \\
               & $a$ & 778.57$\times10^{6}$ & km   & $R_{\rm aj}$ & 11422.8 & r$_J$ \\
               & $e$ &  0.0489 &                   & $q_{\rm tid,pj}$ & -2.82816$\times10^{-9}$ &  \\
               &     &   &                         & $q_{\rm tid,aj}$ & -2.10853$\times10^{-9}$ &  \\
\enddata
\tablenotetext{a}{Physical paramters in left column from \url{ 
https://ssd.jpl.nasa.gov/horizons.cgi}. Derived tidal parameters at perijove and
apjove in right column. }
\label{tab:params}
\end{deluxetable}

%

\floattable
\floattable
\begin{deluxetable}{llrrrrr}
\tabletypesize{\scriptsize}
\tablecolumns{7}
\tablewidth{0pc}
\tablecaption{Interior Models}
\tablehead{
\colhead{Model\tablenotemark{a}} & \colhead{Description} & \colhead{$J_{2}$} & \colhead{$J_{4}$} &  \colhead{$J_{6}$} &  \colhead{$J_{8}$} &  \colhead{$J_{10}$} }
\startdata
Target $J_n$ & PJ06 gravity solution\tablenotemark{b} &  {\bf 14696.51} & {\bf -586.60} & {\bf 34.20} & -2.42 & 0.17 \\
\tableline
A1 & helium rain onset pressure  & 14696.51  & -586.613 & 34.202 & -2.458 & 0.202 \\
A2 & helium rain layer $\Delta S$ & 14696.51 & -586.609 & 34.202 & -2.458 & 0.202 \\
A3 & dilute core $Z$, $r=0.7$  & 14696.51 & -586.602 & 34.204 & -2.457 & 0.202 \\
A4 & dilute core $Z$, $r=0.5$  & 14696.51 & -586.603 & 34.203 & -2.457 & 0.202 \\
\tableline
B1 & \citet{Kaspi2018} longitude independent & 14696.01 & -586.561 & 34.200 & -2.459 & 0.202 \\
B2 & \citet{Kaspi2018} longitude dependent & 14697.04 & -586.463 & 34.200 & -2.460 & 0.202 \\
\tableline
C1 & \citet{Militzer2019b} model 1 & 14694.90 & -586.524 & 34.502 & -2.504 & 0.207 \\
C2 & \citet{Militzer2019b} model 2 & 14695.27 & -586.721 & 34.511 & -2.505 & 0.207 \\
C3 & \citet{Militzer2019b} model 3 & 14694.92 & -586.562 & 34.499 & -2.504 & 0.207 \\
C4 & \citet{Militzer2019b} model 4 & 14695.08 & -586.566 & 34.503 & -2.504 & 0.207 \\
\enddata
\label{tab:models}
\tablenotetext{a}{ All A and B models fit $J_2$ and $J_4$ by tuning $S$ and
$Z2$. Description for A models specifies  parameter tuned to fit $J_6$. B models
are fit to the $\Delta J_n$ from \citet{Kaspi2018} optimized wind profiles.  C
models are more complicated interior profiles from \citet{Militzer2019b}, where
the $\Delta J_n$ correspond to a wind profile optimized simultaneously with
interior density profile.  Model A1 is used as the reference model in all tables
and figures unless otherwise stated.}
\tablenotetext{b}{ Target $J_n$ are from the PJ06 gravity solution \citet{Iess2018} with
the calculated tidal contribution from Io removed. }
\end{deluxetable}

\floattable
\rotate
\begin{deluxetable}{rrllrlllllll}
\tabletypesize{\scriptsize}
\tablecolumns{12}
\tablewidth{0pc}
\tablecaption{Calculated equilibrium Love numbers for Io}
\tablehead{
 $n$ & $m$ &  $k_{nm}$ nonrot &  $k_{nm}$ perijove  & $k_{nm}$
 apojove & $dk_{nm}/dR_{\rm sat}$ &  non-linear & err numerical & err $q_{\rm rot}$ & err interior &
err winds K18 & err winds M20}
\startdata
 2 &  2 &  0.536369 &  0.589759$^{+1.1e-04}_{-1.3e-04}$ &  0.589749 &  -1.931e-04 &        True &  $\pm$4.9e-07 &  $\pm$9.2e-05 &  $^{+0.0e+00}_{-5.7e-06}$ &  $^{+2.1e-05}_{-2.5e-05}$ &  $^{+0.0e+00}_{-3.0e-05}$ \\
 2 &  0 &            &     0.4699$^{+7.1e-04}_{-7.8e-04}$ &     0.4699 &  -2.745e-04 &       False &  $\pm$1.9e-04 &  $\pm$8.4e-05 &  $^{+1.3e-04}_{-4.9e-04}$ &  $^{+2.6e-05}_{-1.5e-05}$ &  $^{+3.1e-04}_{-0.0e+00}$ \\
 3 &  3 &    0.22434 &    0.23948$^{+6.8e-05}_{-6.3e-04}$ &    0.23948 &  -9.256e-05 &        True &  $\pm$1.5e-05 &  $\pm$5.3e-05 &  $^{+0.0e+00}_{-1.9e-05}$ &  $^{+0.0e+00}_{-8.8e-05}$ &  $^{+0.0e+00}_{-5.5e-04}$ \\
 3 &  1 &            &    0.19014$^{+5.7e-05}_{-5.7e-04}$ &    0.19013 &  -1.556e-04 &        True &  $\pm$1.1e-05 &  $\pm$4.5e-05 &  $^{+0.0e+00}_{-1.4e-05}$ &  $^{+0.0e+00}_{-7.0e-05}$ &  $^{+0.0e+00}_{-5.0e-04}$ \\
 4 &  4 &    0.12786 &     0.1353$^{+8.8e-04}_{-8.9e-05}$ &    0.13529 &  -5.881e-05 &        True &  $\pm$1.3e-05 &  $\pm$3.9e-05 &  $^{+5.8e-05}_{-0.0e+00}$ &  $^{+0.0e+00}_{-3.7e-05}$ &  $^{+7.7e-04}_{-0.0e+00}$ \\
 4 &  2 &            &     1.7432$^{+1.7e-03}_{-8.5e-04}$ &     1.7702 &   5.589e-01 &        True &  $\pm$1.1e-04 &  $\pm$2.9e-04 &  $^{+6.2e-08}_{-6.8e-06}$ &  $^{+0.0e+00}_{-4.4e-04}$ &  $^{+1.3e-03}_{-0.0e+00}$ \\
 4 &  0 &            &     1.8231$^{+2.3e-03}_{-2.8e-03}$ &     1.8516 &   5.886e-01 &       False &  $\pm$6.6e-04 &  $\pm$3.4e-04 &  $^{+4.1e-04}_{-1.4e-03}$ &  $^{+0.0e+00}_{-4.5e-04}$ &  $^{+8.8e-04}_{-0.0e+00}$ \\
 5 &  5 &    0.08354 &    0.08806$^{+1.2e-03}_{-5.6e-05}$ &    0.08805 &  -4.136e-05 &        True &  $\pm$1.1e-05 &  $\pm$3.1e-05 &  $^{+9.1e-06}_{-8.1e-07}$ &  $^{+0.0e+00}_{-1.3e-05}$ &  $^{+1.2e-03}_{-0.0e+00}$ \\
 5 &  3 &            &    0.81536$^{+6.7e-03}_{-4.6e-04}$ &    0.82769 &   2.548e-01 &        True &  $\pm$9.1e-05 &  $\pm$1.9e-04 &  $^{+1.1e-04}_{-0.0e+00}$ &  $^{+0.0e+00}_{-1.7e-04}$ &  $^{+6.3e-03}_{-0.0e+00}$ \\
 5 &  1 &            &     0.9406$^{+7.3e-03}_{-5.4e-04}$ &     0.9551 &   2.998e-01 &        True &  $\pm$1.0e-04 &  $\pm$2.4e-04 &  $^{+1.9e-04}_{-0.0e+00}$ &  $^{+0.0e+00}_{-2.1e-04}$ &  $^{+6.7e-03}_{-0.0e+00}$ \\
 6 &  6 &   0.059081 &   0.062151$^{+1.2e-03}_{-1.0e-04}$ &   0.062149 &  -3.100e-05 &        True &  $\pm$9.3e-06 &  $\pm$2.6e-05 &  $^{+0.0e+00}_{-5.8e-05}$ &  $^{+1.3e-05}_{-1.1e-05}$ &  $^{+1.2e-03}_{-0.0e+00}$ \\
 6 &  4 &            &    0.49903$^{+8.0e-03}_{-5.2e-04}$ &    0.50647 &   1.538e-01 &        True &  $\pm$7.0e-05 &  $\pm$1.5e-04 &  $^{+0.0e+00}_{-2.1e-04}$ &  $^{+2.4e-05}_{-8.7e-05}$ &  $^{+7.8e-03}_{-0.0e+00}$ \\
 6 &  2 &            &     5.8999$^{+6.1e-02}_{-3.5e-03}$ &     6.0858 &   3.842e+00 &        True &  $\pm$7.0e-04 &  $\pm$1.0e-03 &  $^{+0.0e+00}_{-7.8e-04}$ &  $^{+0.0e+00}_{-9.6e-04}$ &  $^{+5.9e-02}_{-0.0e+00}$ \\
 6 &  0 &            &      6.826$^{+1.1e-01}_{-1.3e-02}$ &      7.043 &   4.498e+00 &       False &  $\pm$5.6e-03 &  $\pm$1.6e-03 &  $^{+7.3e-04}_{-4.7e-03}$ &  $^{+0.0e+00}_{-1.1e-03}$ &  $^{+1.0e-01}_{-0.0e+00}$ \\
\enddata
\label{tab:Io}
\tablenotetext{a}{See supplementary material for full table in machine readable
    format, \texttt{io\_2019\_12\_02.txt}.}
\tablenotetext{b}{Total error of ``$k_{nm}$ perijove'' includes uncertainties  from colunmns
8-12.}

\end{deluxetable}

\floattable
\begin{deluxetable}{rrllrlllll}
\tabletypesize{\scriptsize}
\tablecolumns{10}
\tablewidth{0pc}
\tablecaption{Calculated equilibrium Love numbers for Europa}
\tablehead{
 $n$ & $m$ &  $k_{nm}$ non-rotating &  $k_{nm}$ perijove  & $k_{nm}$
    apojove & $dk_{nm}/dR_{\rm sat}$ &  non-linear & err numerical & err $q_{\rm rot}$}
\startdata
 2 &  2 &  0.536369 &  0.589414$\pm$9.2e-05 &  0.589408 &  -4.782e-05 &        True &  4.7e-07 &  9.1e-05 \\
 2 &  0 &            &      0.469$\pm$1.3e-03 &      0.469 &  -6.799e-05 &       False &  1.2e-03 &  8.4e-05 \\
 3 &  3 &    0.22434 &    0.23932$\pm$6.9e-05 &    0.23931 &  -2.288e-05 &        True &  1.5e-05 &  5.3e-05 \\
 3 &  1 &            &    0.18986$\pm$5.7e-05 &    0.18985 &  -3.849e-05 &        True &  1.1e-05 &  4.5e-05 \\
 4 &  4 &    0.12786 &    0.13519$\pm$5.3e-05 &     0.13520 &  -1.450e-05 &        True &  1.3e-05 &  3.9e-05 \\
 4 &  2 &            &     4.1975$\pm$9.3e-04 &     4.3662 &   8.893e-01 &        True &  2.6e-04 &  6.7e-04 \\
 4 &  0 &            &      4.407$\pm$9.1e-03 &      4.584 &   9.367e-01 &       False &  8.3e-03 &  8.0e-04 \\
 5 &  5 &   0.083531 &   0.087982$\pm$3.9e-05 &   0.087998 &  -8.086e-06 &        True &  8.3e-06 &  3.1e-05 \\
 5 &  3 &            &     1.9343$\pm$6.7e-04 &     2.0114 &   4.056e-01 &        True &  2.3e-04 &  4.4e-04 \\
 5 &  1 &            &      2.2570$\pm$8.0e-04 &     2.3477 &   4.771e-01 &        True &  2.5e-04 &  5.5e-04 \\
\enddata
\label{tab:Europa}
\tablenotetext{a}{See supplementary material for full table in machine readable
    format, \texttt{europa\_2019\_12\_02.txt}.}
\end{deluxetable}

\floattable
\begin{deluxetable}{rrllrlllll}
\tabletypesize{\scriptsize}
\tablecolumns{10}
\tablewidth{0pc}
\tablecaption{Calculated equilibrium Love numbers for Ganymede}
\tablehead{
 $n$ & $m$ &  $k_{nm}$ non-rotating &  $k_{nm}$ perijove  & $k_{nm}$
    apojove & $dk_{nm}/dR_{\rm sat}$ &  non-linear & err numerical & err $q_{\rm rot}$}
\startdata
 2 &  2 &  0.536369 &  0.589274$\pm$9.2e-05 &  0.589276 &  -1.177e-05 &        True &  4.5e-07 &  9.1e-05 \\
 2 &  0 &            &     0.4692$\pm$1.0e-03 &     0.4692 &  -1.675e-05 &       False &  9.6e-04 &  8.4e-05 \\
 3 &  3 &    0.22434 &    0.23925$\pm$6.9e-05 &    0.23925 &  -5.506e-06 &       False &  1.6e-05 &  5.3e-05 \\
 3 &  1 &            &    0.18975$\pm$5.8e-05 &    0.18975 &  -9.348e-06 &       False &  1.3e-05 &  4.5e-05 \\
 4 &  4 &    0.12786 &    0.13515$\pm$5.3e-05 &    0.13516 &  -2.322e-06 &        True &  1.3e-05 &  3.9e-05 \\
 4 &  2 &            &    10.7058$\pm$2.3e-03 &    10.7315 &   1.418e+00 &        True &  6.7e-04 &  1.7e-03 \\
 4 &  0 &            &      11.26$\pm$1.7e-02 &      11.29 &   1.494e+00 &       False &  1.5e-02 &  2.0e-03 \\
 5 &  5 &   0.083498 &   0.087961$\pm$3.3e-05 &   0.087978 &  -4.765e-05 &        True &  2.1e-06 &  3.1e-05 \\
 5 &  3 &            &     4.9016$\pm$1.9e-03 &     4.9138 &   6.467e-01 &       False &  7.5e-04 &  1.1e-03 \\
 5 &  1 &            &      5.7480 $\pm$2.1e-03 &     5.7622 &   7.608e-01 &       False &  7.4e-04 &  1.4e-03 \\
\enddata
\label{tab:Ganymede}
\tablenotetext{a}{See supplementary material for full table in machine readable
    format, \texttt{ganymede\_2019\_12\_02.txt}.}
\end{deluxetable}

\floattable
\begin{deluxetable}{rrllrlllll}
\tabletypesize{\scriptsize}
\tablecolumns{10}
\tablewidth{0pc}
\tablecaption{Calculated equilibrium Love numbers for Callisto}
\tablehead{
 $n$ & $m$ &  $k_{nm}$ non-rotating &  $k_{nm}$ perijove  & $k_{nm}$
    apojove & $dk_{nm}/dR_{\rm sat}$ &  non-linear & err numerical & err $q_{\rm rot}$}
\startdata
 2 &  2 &  0.536369 &  0.589214$\pm$9.8e-05 &  0.589218  &  -2.165e-06 &       False &  5.0e-07 &  9.8e-05 \\
 2 &  0 &            &      0.469$\pm$9.6e-03 &       0.47 &  -3.073e-06 &       False &  9.5e-03 &  1.0e-04 \\
 3 &  3 &    0.22434 &    0.23922$\pm$4.8e-05 &    0.23925 &  -1.060e-06 &       False &  2.8e-05 &  2.0e-05 \\
 3 &  1 &            &     0.18970$\pm$4.1e-05 &    0.18972 &  -1.702e-06 &       False &  2.1e-05 &  2.0e-05 \\
 4 &  4 &    0.12776 &    0.13513$\pm$1.6e-05 &    0.13518 &  -1.662e-06 &        True &  1.1e-05 &  4.7e-06 \\
 4 &  2 &            &     32.507$\pm$2.2e-03 &     33.434 &   2.496e+00 &        True &  2.0e-03 &  1.6e-04 \\
 4 &  0 &            &       34.2$\pm$4.9e-01 &       35.2 &   2.628e+00 &       False &  4.9e-01 &  1.6e-03 \\
\enddata
\label{tab:Callisto}
\tablenotetext{a}{See supplementary material for full table in machine readable
    format, \texttt{callisto\_2019\_12\_02.txt}.}
\end{deluxetable}

\floattable
\begin{deluxetable}{rrllrlllll}
\tabletypesize{\scriptsize}
\tablecolumns{10}
\tablewidth{0pc}
\tablecaption{Calculated equilibrium Love numbers for Solar tide}
\tablehead{
 $n$ & $m$ &  $k_{nm}$ non-rotating &  $k_{nm}$ perijove  & $k_{nm}$
    apojove & $dk_{nm}/dR_{\rm sat}$ &  non-linear & err numerical & err $q_{\rm rot}$}
\startdata
 2 &  2 &   0.536369 &  0.589186$\pm$9.1e-05 &  0.589188 &   1.878e-12 &       False &  4.6e-08 &  9.1e-05 \\
 2 &  0 &            &       0.469$\pm$1.3e-03 &       0.469 &  -3.756e-12 &       False &  1.2e-03 &  8.4e-05 \\
\enddata
\label{tab:Solar}
\tablenotetext{a}{See supplementary material for full table in machine readable
    format, \texttt{solar\_2019\_12\_02.txt}.}
\end{deluxetable}



\end{document}